\documentclass[9pt,twocolumn,twoside]{pnas-new}
\usepackage{graphics,color,graphicx,amsmath}
\usepackage{xr}
\usepackage{booktabs}

\newcommand{\fancyP}{\mathcal{P}}
\newcommand{\fancyR}{\mathcal{R}}

\newcommand{\br}[1]{\left\langle#1\right\rangle}
\newcommand{\session}[1]{$#1^{\rm th}$}
\newcommand{\sessionn}[1]{$#1^{\rm nd}$}

\templatetype{pnasresearcharticle} 

\title{Collective contributions to polarized voting}
\author{Edward D.~Lee}
\affil{Department of Physics \& Astronomy, Seoul National University, 08826 Seoul, Republic of Korea\\
Complexity Science Hub, Metternichgasse 8, 1030, Vienna, Austria}
\dates{This manuscript was compiled on \today}

\leadauthor{Lee} 

\significancestatement{Discussion of politics today often revolves around a partisan contest, which is reflected in models. Yet, voting in political bodies is more complex. Elaborating on an influential physics model for machine learning, we accurately model complex voting patterns by combining voters' preferences and the shared context of the vote (such as the issues at play or the voting rules). The second, collective aspect predominantly explains increased polarization in the U.S.~Senate. This breaks with the unidimensional and zero-sum game approach to ideological polarization and brings out the importance of collective choices for bridging the political divide.}

\authordeclaration{The authors declare no other competing interests.}
\correspondingauthor{\textsuperscript{2}To whom correspondence should be addressed. E-mail: eddielee@snu.ac.kr}


\begin{abstract}
The crux of polarized voting in institutions like legislatures or courts is separation along a unidimensional ideological axis, but voting behavior is in reality more complex, with other signatures of collective order. We extend a foundational, statistical physics framework, restricted Boltzmann machines, to capture both such aspects of voting. The models are minimal, fit strongly correlated voting data, and have parameters that transparently give vote probabilities. The model accounts for multi-dimensional voter preferences and the context in which such preferences are expressed to disentangle individual from collective contributions; for example, legislative bills can negotiate multiple issues, whose appeal adds up or competes for individual votes. With the example of the U.S.~Senate, we find that senators have multi-dimensional preferences, and, as one consequence, non-polarized coalitions coexist with polarized ones. Increasing polarization is predominantly explained by fewer votes that elicit bipartisan coalitions. We show that these accounts can be consistent, if far more parsimonious, than interaction-driven order. The findings highlight the collective choice of the content and the rules of voting in the ebb and flow of polarization.
\end{abstract}

\begin{document}
\maketitle

\thispagestyle{firststyle}
\ifthenelse{\boolean{shortarticle}}{\ifthenelse{\boolean{singlecolumn}}{\abscontentformatted}{\abscontent}}{}

\dropcap{P}olarization is a widely discussed issue today, and it is especially concerning when considering its repercussions on political compromise, consensus, and the design of effective policy \cite{baronchelliEmergenceConsensus2018, iyengarOriginsConsequences2019, axelrodPreventingExtreme2021}. An example is the deepening two-party, blue vs.~red, liberal vs.~conservative divide in the U.S. \cite{masudaCanPartisan2011, palDepolarizationOpinions2023}. In the electorate, apparent signs of polarization reflect in part better voter sorting into parties, and not necessarily increasing extremism at the level of the individual \cite{fiorinaPoliticalPolarization2008, baldassarriPartisansConstraint2008, yangWhyAre2020, thurnerWhyMore2025}. The evidence for hardening partisan divisions is stronger, if not quite ubiquitous, in political institutions such as Congress \cite{hillDisconnectRepresentation2015} and the Supreme Court \cite{martinDynamicIdeal2002}, where political actors adhere more closely to party lines and have become more antagonistic in their policy agendas \cite{theriaultPartyPolarization2008}. Supporting such political analyses are voting preference models \cite{pooleSpatialModel1985}, which return scalar ideology scores on which the political parties increasingly diverge \cite{mccartyDefenseDWNOMINATE2016}. Popular and scholarly accounts of a growing political divide indicate that polarization is strengthening, especially in terms of how it is expressed in polarized voting, and as a corollary that polarized voting patterns are a sufficient explanation of voter behavior.

Yet, there are a few problems with a focus on the binary, partisan aspect of voting. For one, it overlooks the complexity of empirical voting patterns, including many ``non-ideological'' examples \cite{sirovichPatternAnalysis2003, lawsonSpectralAnalysis2006, giansiracusaComputationalGeometry2019}. Non-partisan voting, including consensus and other split votes that do not align with partisan expectation are common even in polarized systems. The U.S.~Supreme Court frequently finds itself in strong consensus --- 30-40\% of recorded votes are consistently unanimous or with a single defection since 1946 \cite{spaethSupremeCourt2016} --- and displays many divided but ``surprising'' non-partisan voting configurations, whose frequencies resemble a heavy-tailed Zipf's law \cite{leeStatisticalMechanics2015}. The U.S.~Congress also displays consensus on substantive issues, despite popular perception (if more often in the Senate than in the House of Representatives), examples of which are often overlooked in the analysis of voting models or excluded from the data \cite{pooleScalingRoll2011}. Thus, a more accurate pronouncement might be that there are contexts in which polarization is expressed and others in which it is not because votes depend on the options in question. If the choice of what to vote on is as important as the vote, this is necessarily excluded from analysis that focuses on a single ideological dimension \cite{martinDynamicIdeal2002} or is not explicitly modeled when accounting for interactions \cite{guimeraJusticeBlocks2011, leePartisanIntuition2018, leeStatisticalMechanics2015}. Yet, the notion that individuals simply express partisan or ideological preferences emphasizes their individual motivations but not the fact that those motivations may reflect \textit{interactions} with colleagues \cite{cliffordModelSpatial1973, hagleFreshmanEffects1993, maltzmanStrategicPolicy1996}. 
In short, a partisan account misses the substantial amount of statistical information in the tail of the voting distribution, which could be either a telltale signal of voting context or interactions. 
We develop a model that can capture the distribution of votes, in contrast to existing models that cannot, and thus can show that political voters in the U.S.~Senate operate in a multi-dimensional space, and this allows for multiple coalitions to coexist, bipartisan and the many ways to be partisan. Furthermore, this formulation allows us to show how it, and variations thereof, are analytically equivalent to classes of interaction models \cite{kruisThreeRepresentations2016, leeValenceInteractions2024}, unifying two, seemingly, disparate approaches to political voting. 

We do so by formulating a statistical physics model that addresses these shortcomings while embodying a central tenet of voting theory. The voters in the model have multi-dimensional internal preferences that become relevant depending on the ``issues'' at play in any particular vote. We show that this is a generalization of the restricted Boltzmann machine (RBM) \cite{hintonBoltzmannMachines1984}. Besides capturing complex features like the distribution of consensus accurately, it accomplishes this with many fewer parameters that would be necessary for spatial voting models of comparable performance \cite{pooleSpatialModel1985, pooleSpatialModels2005}.

\begin{figure}
	\centering
	\includegraphics[width=\linewidth]{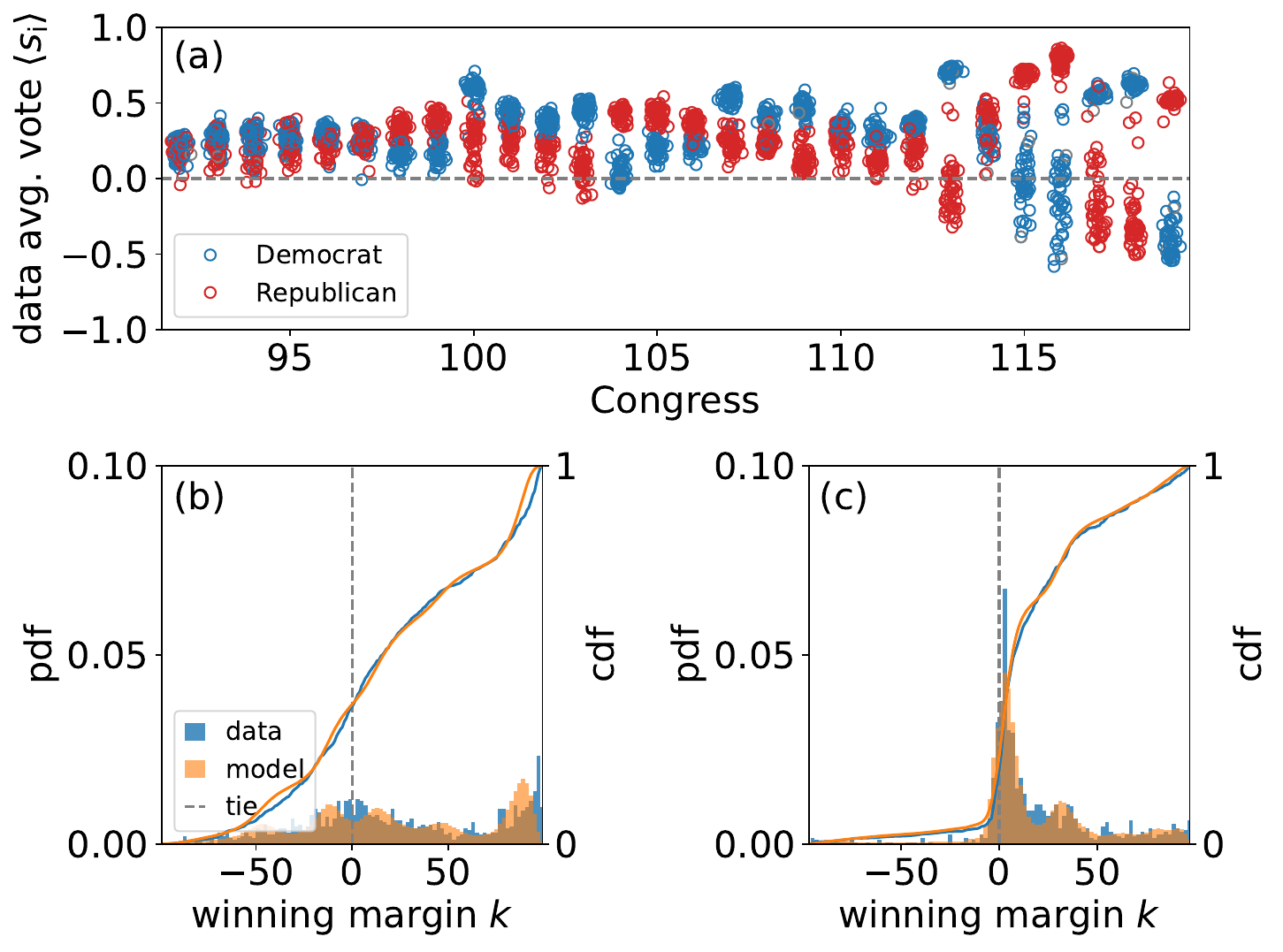}
	\caption{(a) Average `yea' or `nay' vote over time by Democratic (blue) and Republican (red) senators in the modern era of the U.S.~Senate. The majority party votes `yea' more often than the minority as would be expected in a majority-rule body. Distribution of votes with a winning margin of $k$ for (b) the \session{97} Congress (1981-1983) and (c) the \session{117} Congress (2021-2023). Model with $K=4$ is overlaid on the data in orange. It is clear that in the \session{117} Congress (2021-2023) that many more votes occur right at the boundary of the majority and a substantially small fraction of votes are passing with near unanimous support. While this is an indication of polarization, it is also the case that a substantial fraction of votes ($\small\sim$50\%) do not fall along the expected 50-50 divide from strict party-line voting. }\label{gr:vote margin}
\end{figure}

We study roll-call voting from two-year sessions of the US Senate. The Senate is one of two houses in Congress, which is the federal legislature. Congress is responsible for creating, debating, and passing laws that govern the nation. The Senate is composed of $N=100$ members (with small, transient fluctuations below this number in the time period that we consider), with two senators representing each state. A rotating third of senators undergo an election on a biennial basis, and senators are elected to serve terms of six years. The modern era of legislative voting --- where reforms (e.g.~Legislative Reorganization Act of 1970) dramatically increased the number of recorded votes --- consists of the last five decades (1971-2025), and for this period we draw data from the Voteview project \cite{lewisVoteviewCongressional2025}. Of all votes conducted in the U.S. Senate, roll-call votes are the only category  in which each senator's vote is recorded. Formally, each roll-call vote is representable by a vector $\vec{s} \in \{-1, 1\}^N$, listing for each  voter i, the vote for approval $s_{\rm i} = 1$ (`Yea'), for rejection $s_{\rm i} = -1$ (`Nay'), and a missing entry for absence and abstention. The outcome, passage or rejection, is generally determined by a simple majority of present voters (the quorum) except for particular votes that require a 2/3 or 3/5 majority. We treat absent votes as missing data, in which one might presume a vote could have been measured had the Senator been queried in accordance with the rest of their voting record. In total, each session comprises of approximately $R\sim10^3$ roll-call votes, of which each Senator will on average vote in ${\small\sim}950$ (more details in Materials \& Methods). Thus, the data is sparse relative to the number of possible voting outcomes, $2^N$, posing a challenge for learning highly parameterized models because number of observed data points limits the complexity of models that can be inferred.

As an overview of the divergence in Democratic vs.~Republican voting patterns over the last few decades, we show the average vote of each senator indexed i in Figure~\ref{gr:vote margin}a,
\begin{align}
	\br{s_{\rm i}} &\equiv \sum_{\vec s} P(\vec s)s_{\rm i},\label{eq:si}
\end{align}
where each vote occurs with a probability $P(\vec s)$. By coloring each senator by their respective party (blue for Democrat and red for Republican), we can make out the winning majority party by a bias towards `yea.' It is also clear that the partisan gap has increased in recent congressional sessions, leading to both a wider spread in the average vote between the parties and even a discernible gap between the blue and red points. This is not a perfect trend, and we see, for example, that the \session{100} Congress has an unusually wide spread for the era and, after the first major deviation in the \session{113} Congress, the \session{114} Congress converges.\footnote{This was during the last two years of Obama's presidency, when the Republicans took both the House and the Senate, the first such occurrence in a decade.} As a closer look at the partisan gap, we show in Figure~\ref{gr:vote margin}b and c the observed frequency of the winning margins in the \session{97} and \session{117} Congresses $\hat p(k=\sum_{\rm i=1}^N s_{\rm i})$. The peak around $k=0$ indicates the importance of the majority cutoff in a closely divided body. Despite the popular focus on partisan divisions, the second peak near $k\geq 67$ for the \session{97} Congress indicates a large fraction 41\% of strongly bipartisan votes (and 13\% for bipartisan nay votes). While the \session{117} Congress in panel c displays noticeably less agreement, bipartisan yea votes still constitute a substantial 27\% of votes (with 3\% bipartisan nays). Thus, the observed margin distribution $\hat p(k)$ reveals a strong consensus force at odds with a partisan story.

\begin{figure}[t!]
    \centering
    \includegraphics[width=.49\linewidth]{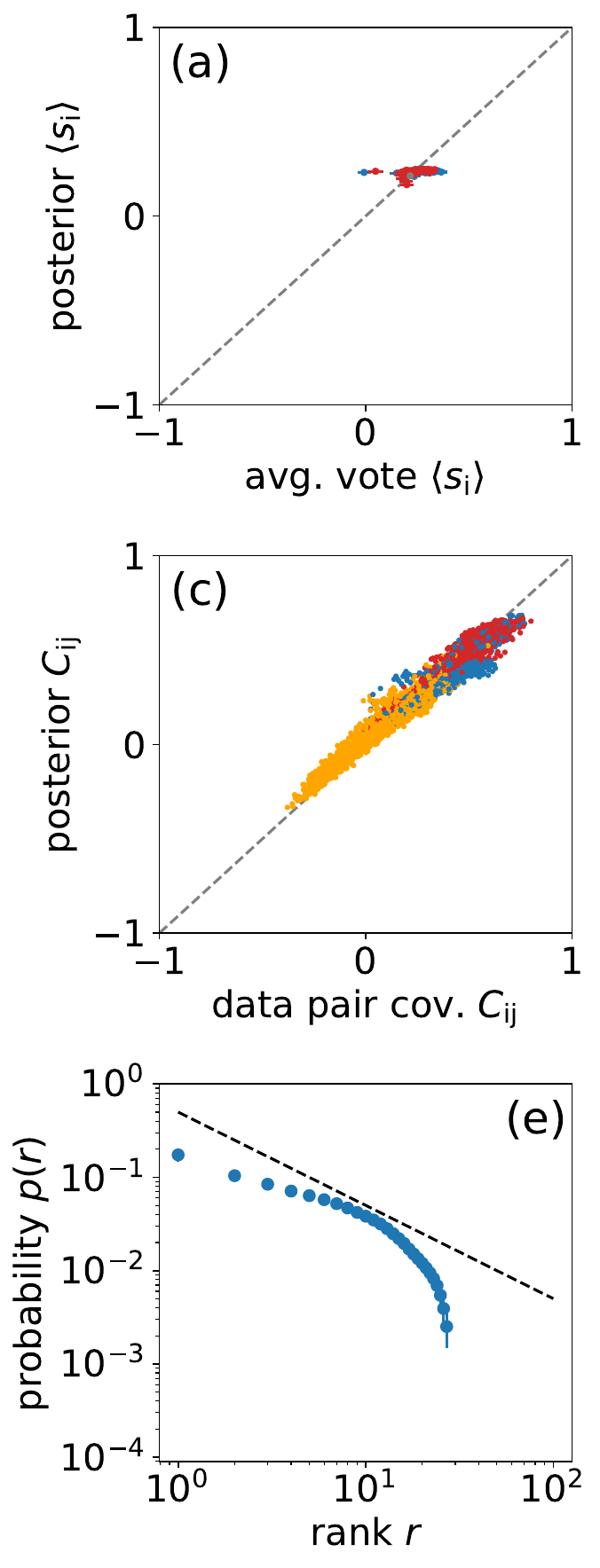}
    \includegraphics[width=.49\linewidth]{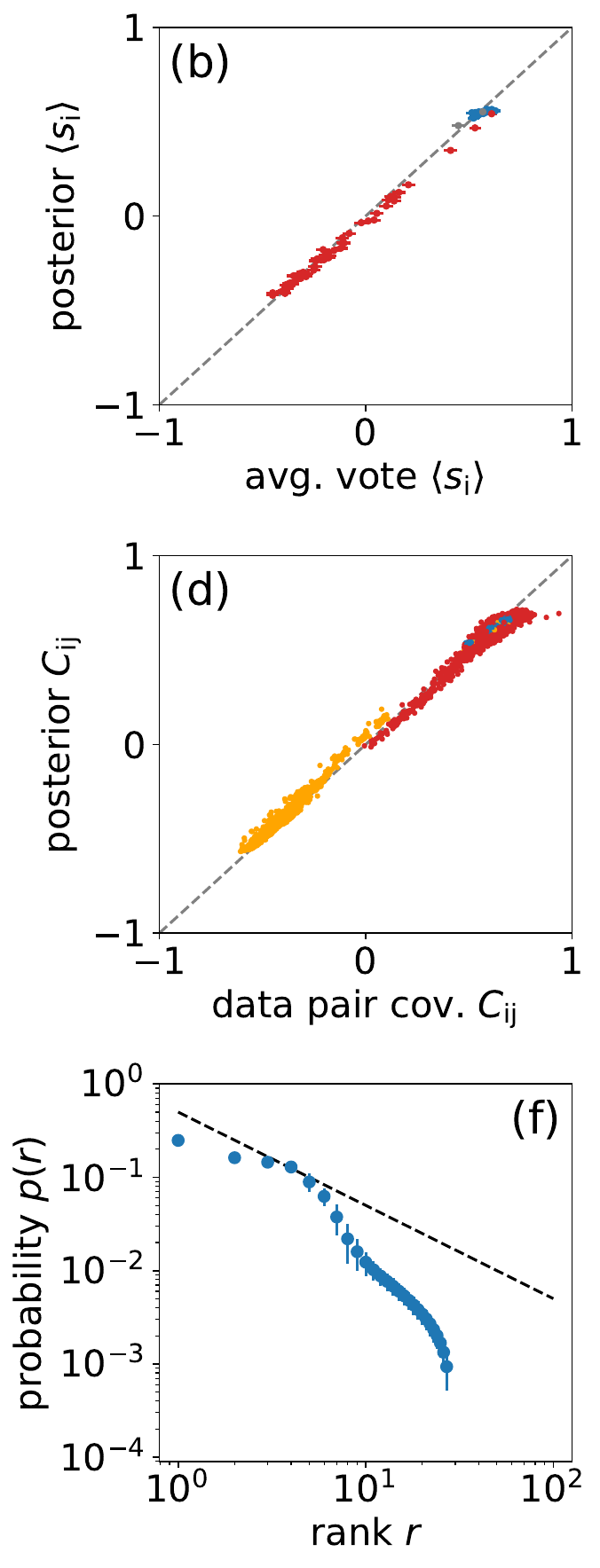}\\
    \caption{Model comparison with data for \session{97} and \session{117} Congresses and $K=3$. (a, b) Average votes, $\langle s_{\rm i} \rangle_{\text{data}}$ and (c, d) pair covariances $C_{\rm ij}$. (e, f) Zipf's plot for inferred context probability distribution $P(\vec\sigma)$ when ordered by probability rank $r$. Dashed, black line is Zipf's law $p(r)\propto r^{-1}$ for comparison. All error bars show a standard error over the posterior sample.}
    \label{fig:observables}
\end{figure}

The collective voting patterns in the Senate contain both bipartisan and competitive coalitions, which belie a singular focus on two-party competition. Consider the pairwise covariance in Figure~\ref{fig:observables},
\begin{align}
	C_{\rm ij} &\equiv \br{s_{\rm i}s_{\rm j}} - \br{s_{\rm i}}\br{s_{\rm j}}.\label{eq:sisj}
\end{align}
We color the covariance between Democrats (blue), Republicans (red), and between the two parties (orange).  A simple block model of parties, without individual or pairwise heterogeneity, would show similar rate of agreement for all voters of the same party within their respective party and against the opposing party, but instead we see that the inter-party covariance, where one voter is Democrat and the other Republican, span a wide range. This is especially noticeable for the \session{97} Congress, where the voter averages are tightly distributed around $\br{s_{\rm i}}\approx1/4$, but the covariance even within the party blocs can range from $0\lesssim \br{s_{\rm i}s_{\rm j}} \lesssim 1$.\footnote{This is not to say that the partisan divides are unimportant. As we see, the inter-party covariance shift from overwhelmingly positive to negative between the two sessions.} The \session{117} shows an even wider spread in the intra-party covariances. The spread indicates that individual senators' typical voting patterns, while often aligning with partisan expectations, obscure enormous variation in how they align with or against their party across different votes.\footnote{As a corollary, the two peaks in the distribution of majority margin $p(k)$ are more complicated than just partisan split and bipartisan rapprochement since many different subsets of voters including partisan and non-partisan coalitions can defect.} That the correlation structure is nontrivial means that a rich and heterogeneous distribution of patterns hide behind the partisan divide.

\section{Model definition}
To model the voting structure, we borrow a key tenet in political science: voters have preferences that bias the probability of a vote in either direction of yea or nay \cite{pooleSpatialModel1985, segalSpatialModel1992, segalSupremeCourt2002, martinDynamicIdeal2002}. The simplest statistical voting model that endows each voter with a bias is the product of independent voters \cite{leeStatisticalMechanics2015, leeValenceInteractions2024}. Denoting the independent model $P_I$, the probability that a randomly chosen roll-call vote yields the vector of votes $\vec{s}$ is
\begin{align}
\begin{aligned}
    P_I(\vec{s}) &=  \frac{e^{-H_I(\vec s)}}{\prod_{\rm i=1}^N e^{h_{\rm i}} + e^{-h_{\rm i}}}\\
    H_I(\vec s) &= -\sum_{\rm i=1}^N h_{\rm i} s_{\rm i} 
\end{aligned}\label{eq:ind1}
\end{align}
where we have written each voter's tendency in terms of a bias, or field, $h_{\rm i}$, with the normalization term $e^{h_{\rm i}}+e^{-h_{\rm i}}$. In statistical physics, $H_I(\vec s)$ is known as the Hamiltonian, or energy function. The field $h_{\rm i}$ indicates the tendency with which the voter i tends to vote `yea' when $h_{\rm i}>0$ or `nay' when $h_{\rm i}<0$. When $h_{\rm i}=0$, the probability of voting in either direction is equal. The independent model in Eq~\ref{eq:ind1} presents a way of capturing exactly the expected vote of each voter $\br{s_{\rm i}}$, but it would fail to capture any additional higher-order structure in the probability distribution.

In particular, Eq~\ref{eq:ind1} fails to capture how voters might consider the contents of the bill at question, which determines the framing of the vote (e.g.~voting ``yea'' on a pro-abortion rights bill would be like voting ``nay'' on an anti-abortion rights bill \cite{leeValenceInteractions2024}). We capture this with a contextual cue $\sigma\in\{-1,0,1\}$ such that the product $\sigma h_{\rm i}$ ensures that the preference aligns with the contextual framing of the issue; when the context is irrelevant, $\sigma=0$. By summing over the three possible values of $\sigma$, each occurring with probability $Q(\sigma)$, we have the \textit{single-issue model}
\begin{align}
    P_1(\vec{s}) &= \sum_{\sigma=-1}^1 Q(\sigma) \prod_{\rm i=1}^N \frac{e^{\sigma h_{\rm i} s_{\rm i}}}{e^{h_{\rm i}} + e^{-h_{\rm i}}}.\label{eq:ind2}
\end{align}
Eq~\ref{eq:ind2} corresponds to an extension of the independent voter model in Eq~\ref{eq:ind1}, where the voters remain independent once conditioned on contextual issue $\sigma$, but its presence induces correlations in their votes.

The single-issue model would capture a unidimensional preference, like the liberal-conservative dimension, but voters in reality have multiple, potentially competing preferences \cite{heiderAttitudesCognitive1946, medoFragilityOpinion2021}. In order to capture this aspect, we generalize the fields and contextual cues to be multi-dimensional with $K$ dimensions, representing $K$ different tendencies. Since preferences may align constructively, destructively, or be irrelevant for a particular vote, the effective bias for voter i is given by a linear combination $h_{\rm i}^{\rm eff} = \vec{\sigma}\cdot\vec{h}_{\rm i}$, or a projection onto the shared context or issue space defined by the ternary context vector $\vec{\sigma} \in \{-1,0,1\}^K$, which can take any of $3^K$ values. Its values might represent how multiple issues are framed in a bill in order to produce a coalition. For any single context $\vec\sigma$, we then have the Hamiltonian
\begin{align}
	H(\vec{\sigma},\vec{s}) &= -\sum_{\rm i=1}^N h_{\rm i}^{\rm eff}(\vec\sigma) s_{\rm i},
\end{align}
which implicitly depends on $K$ through $\vec\sigma$. If we do not know \textit{a priori} how often any particular context $\vec\sigma$ occurs, we should marginalize over all the possible contexts $\vec\sigma$ to obtain the observed distribution $P(\vec{s}) = \sum_{\vec{\sigma}} P(\vec{s}|\vec{\sigma}) Q(\vec{\sigma})$. Assuming that we are completely na\"{i}ve with respect to the probability of any single $\vec\sigma$ such that $P(\vec\sigma)\propto e^{-g(\vec\sigma)}$ for some contextual ``energy'' function $g(\vec\sigma)$, we obtain the $K$-issue model,
\begin{align}
	P_K(\vec{s}) &= \frac{1}{Z_K}\sum_{\vec\sigma\in\{-1,0,1\}^K} e^{-H(\vec{\sigma},\vec{s})-g(\vec\sigma)}\label{eq:K model}
\end{align}
for normalization $Z_K= \sum_{\vec\sigma} e^{-g(\vec\sigma)} \left[\prod_{\rm i=1}^N (e^{h_{\rm i}^{\rm eff}(\vec{\sigma})} + e^{-h^{\rm eff}_{\rm i}(\vec{\sigma})})\right]$. Eq~\ref{eq:K model} presents a powerful generalization of the conditionally independent voter model, where the joint distribution of voters displays nontrivial correlations of all orders with interpretable latent variables that directly represent relative frequencies of the political issues at play.\footnote{Note that the unidimensional $K=1$ model from Eq~\ref{eq:ind2} lies in a subspace of the full model, meaning that we could in principle recover it from the data even after allowing for $K>1$. }

It is worthwhile to note the similarities (and differences) between Eq~\ref{eq:K model} and the maximum entropy, or maxent, approach, which has been used to model collective voting outcomes in the U.S.~Supreme Court \cite{leePartisanIntuition2018, leeStatisticalMechanics2015, leeValenceInteractions2024}. In the maxent approach, the goal is to build a model with the simplest statistical structure given relevant constraints. Using the unique measure of the randomness in a distribution under a few minimal conditions, the Shannon entropy defined as $S[P] = -\sum_{\vec s} P(\vec s)\log P(\vec s)$ \cite{shannonMathematicalTheory1948}, we maximize the entropy while constraining the average vote (Eq~\ref{eq:si}) and pairwise correlations (Eq~\ref{eq:sisj}) to obtain the probability distribution
\begin{align}
	P_{\rm Is}(\vec s) &= \frac{1}{Z_{\rm Is}}e^{\sum_{\rm i} h_{\rm i}s_{\rm i} + \sum_{\rm i<j} J_{\rm ij}s_{\rm i} s_{\rm j}}.\label{eq:ising}
\end{align}
Eq~\ref{eq:ising} is the Ising model, or pairwise maxent model, in statistical physics \cite{jaynesInformationTheory1957}. As with Eq~\ref{eq:K model}, the biases $h_{\rm i}$ represent the voter preference, or bias, but additionally the couplings $J_{\rm ij}$ capture the tendency of voters i and j to agree when $J_{\rm ij}>0$ or to disagree when $J_{\rm ij}<0$. The normalization constant is $Z_{\rm Is}$. 
While Eq~\ref{eq:ising} represents all higher-order marginals as the product of pairwise interactions terms $P_{\rm Is}(\vec s) \propto \prod_{\rm i<j} \phi_{\rm ij}(s_{\rm i}s_{\rm j}) \prod_{\rm i} \phi_{\rm i}(s_{\rm i})$, the $K$-issue model in Eq~\ref{eq:K model} is not factorizable in such a way. 

This is exactly the idea behind restricted Boltzmann machines, composed of two layers: hidden and visible units, where the visible units represent the senators' votes. The hidden units facilitate indirect couplings between the visible units \cite{hintonBoltzmannMachines1984}. As another crucial difference from the Ising model, Eq~\ref{eq:K model} can, with a judicious choice of $K$, present a simpler model with $3^K+N$ parameters instead of $\binom{N}{2}\sim N^2/2$, which for the Senate considerably surpasses the data size because $N^2/2 = 5,000 >10^3$. Thus, Eq~\ref{eq:K model} presents a model that is motivated from a central tenet of voter behavior \cite{segalSupremeCourt2002} and is able to capture higher-order correlations of all orders without a concomitant cost in parameterization that the maxent approach entails \cite{berettaStochasticComplexity2018}.

To solve for the model parameters, we would like to fit the statistics of the roll-call votes $\{\vec x\}$. One way to do this is to maximize the chances that the data are produced by the model, which corresponds to maximizing the likelihood $p(\{\vec x\}|\theta)$ for parameters $\theta\equiv (\{h_{\rm i}^{\rm eff}\}, \{g(\vec\sigma)\})$. Maximum likelihood, however, is prone to overfitting because it ignores the fact that one must also specify the model parameters uniquely from a much wider set. Instead, Bayes' Theorem provides a way to obtain the posterior probability and furthermore encodes the cost of specifying the parameters in the prior \cite{mackayInformationTheory2003},
\begin{align}
	p(\theta | \{\vec x\}) &= \frac{p(\{\vec x\}|\theta)p(\theta)}{p(\{\vec x\})}.\label{eq:posterior}
\end{align}
When the prior $p(\theta)$ is thin, we must pay a large cost for specifying tightly the parameters, and when it is broad, we pay less of a cost, formalizing Occam's Razor. In principle, we can specify the prior, calculate the likelihood, and then use Monte Carlo sampling to sample from the posterior, identifying the most likely outcomes (see Materials \& Methods for more details).

\begin{figure}[t!]
\centering
	\includegraphics[width=\linewidth]{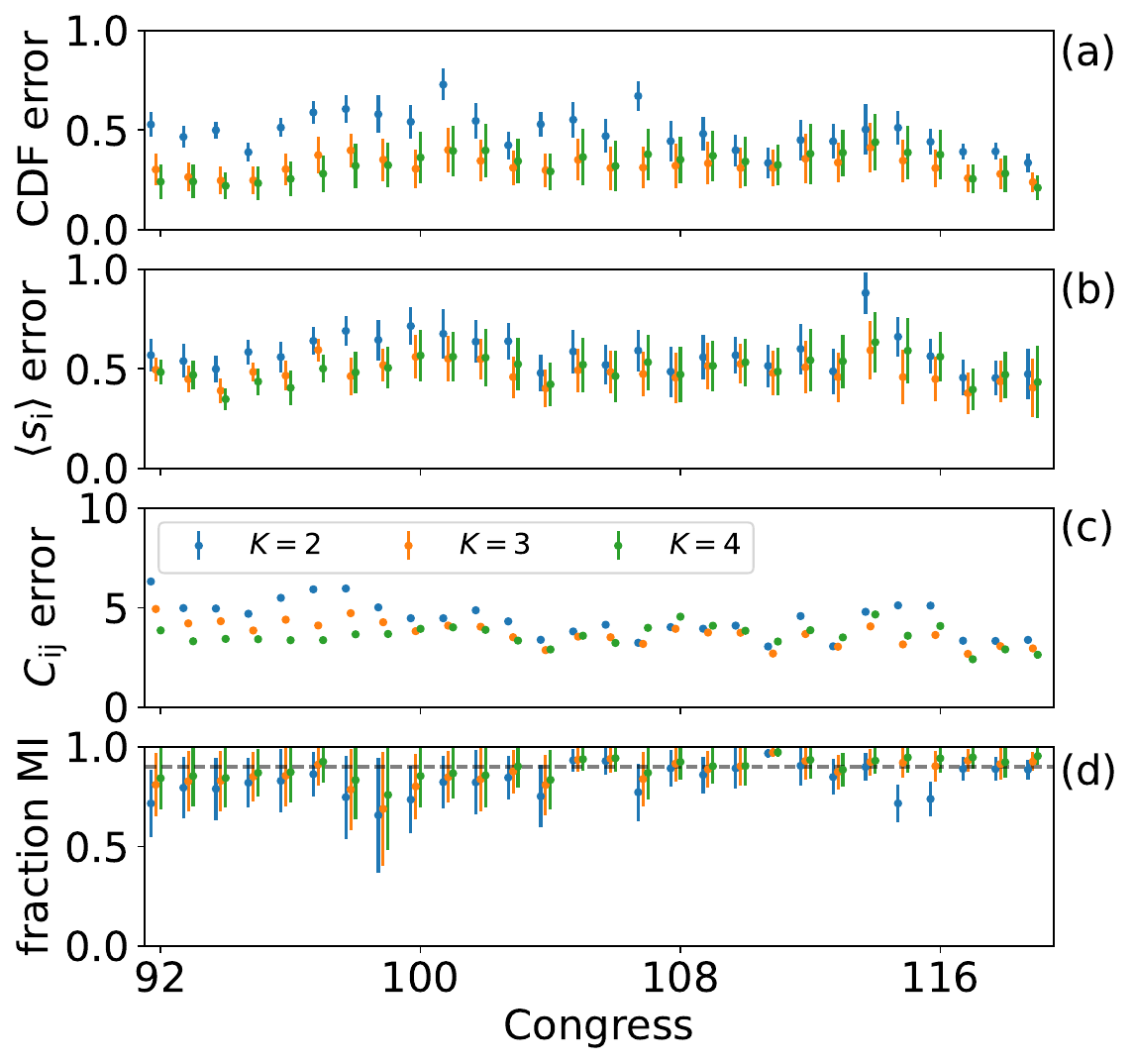}
	\caption{Model performance over congressional sessions. (a) Error on CDF over averaged distribution of winning margin $p(k)$. Root sum of squared errors over (b) voter averages $\br{s_{\rm i}}$ and (c) correlations $C_{\rm ij}$ averaged over posterior sample. Error bars represent a standard error over the posterior sample. (d) Fit to $N=11$ median voters as measured by the fraction of multi-information captured (Eq~\ref{eq:multi-info}). Error bars are the standard deviation over the posterior sample.}\label{gr:fit}
\end{figure}

\section{Incorporating consensus}
At this point, the fields $\vec h_{\rm i}$ are completely unconstrained, but this leaves out the fact that some votes are much more likely to invoke bipartisan support. Inspired by previous work that proposed a way to account for consensus \cite{leeValenceInteractions2024}, we add an additional constraint to the fields: we enforce one field dimension to be uniform across the voters. The shared force can swing the entire system towards yea or nay,
\begin{align}
	h_{\rm unan} &= (\vec h_{\rm i})_{\rm k} = (\vec h_{\rm j})_{\rm k},
\end{align}
for a fixed index k to a dimension of the fields. The consensus field substantially reduces the parameter space and concomitant degeneracies, enforces a kind of minimality, consistent with cases of shared, bipartisan agreement, that ultimately improves model performance.

\section{$K=3$ context model performs well, $K=4$ better}
The dimensionality of the voter fields $K$, including one consensus field, is a key parameter. Increasing $K$ would allow us to obtain an increasingly refined model of voter behavior, but it would come at the cost of specifying more parameters than could be reasonably supported by the finite roll-call data. 
As a measure of performance over the collective statistics, we compare the difference between the distributions over the cumulative winning margin $\sum_{k'=0}^n p(k')$ averaged over the posterior against that from the model. The $K=3$ model returns precise fits to the cumulative distribution, with little improvement for $K=4$. 
A similar pattern echoes in the errors on the average vote $\br{s_{\rm i}}$ and pairwise covariances $C_{\rm ij}$, which converge for $K=3$, with small improvements for $K=4$ in earlier sessions compared to later sessions in Figures~\ref{gr:fit}c and d. As the examples in Figure~\ref{fig:observables} show, the posterior sample captures excellently the mean and pairwise correlations even for the \session{97} (panels a and c) and \session{117} (panels b and d) Congresses. Thus, the $K=3$ model captures the overall patterns of consensus and division, with small additional gains to be made for the earlier sessions by increasing $K$.

The most comprehensive test of the model would be to compare the observed distribution of votes $p(s)$ with the model, but a direct comparison is impossible with the size of the system and missing votes. A difficult yet tractable test uses the subset of the most ``unpredictable,'' median voters, the voters who sit between the most positive and negative values of $\br{s_{\rm i}}$. These would constitute, in roughly equally divided Senates, the most unpredictable members. Choosing a median group of size $n$, we can obtain an estimate of the probability distribution $Q(\vec s)$ of votes where they all vote together and compare it with the model's predictions. The Kullback-Leibler (KL) divergence between the data probability distribution $Q(\vec s)$ and the model $P(\vec s)$ gives such a quantity,
\begin{align}
	D_{\rm KL}[Q||P_K] &= \sum_{\vec s} Q(\vec s)\log\left( \frac{Q(\vec s)}{P(\vec s)} \right),\\
		&= H[Q,P]-H[Q].\label{eq:dkl2}
\end{align}
The KL divergence is semi-positive and unbounded, but we might compare it with the na\"ive independent voter model (Eq~\ref{eq:ind1}) as a baseline.\footnote{With a finite sample, estimation of the entropy $H[q]$ is nontrivial because the na\"ive estimator suffers from bias, but not the cross entropy $H[q,p_K]$. We use the NSB estimator to obtain an estimate of the entropy in Eq~\ref{eq:dkl2}.} To normalize the measure of performance, we take the ratio,\footnote{Note that the denominator decomposes to $H[q]-H[p_0]$, the difference between the entropies of the data and the independent voter model, where we again employ the NSB estimator.}
\begin{align}
	MI &= 1- \frac{D_{\rm KL}[Q||P_K]}{D_{\rm KL}[Q||P_0]}. \label{eq:multi-info}
\end{align}
Eq~\ref{eq:multi-info} is a variation on the multi-information captured and indicates, when $MI=0$, performance on par with the independent model and, when $MI=1$, a perfect match to the observed probability distribution \cite{schneidmanWeakPairwise2006}.

As we show in Figure~\ref{gr:fit}, the $K=3$ model performs well for groups of median voters $N=11$, similarly to $K=4$ (other $N$ shown in ``Supplementary Information''). For the \session{105} sessions and beyond, it captures about 90\% of the multi-information. For earlier sessions, the model does almost as well with smaller group sizes, but the larger groups fall to around 70-80\% of MI captured. One explanation of this would be that earlier sessions of Congress were less polarized as we have seen earlier and thus more complicated. On the other hand, it is also the case that earlier sessions are less well-sampled (relative to their entropies), so the performance is not largely different between the two eras within error bars. Overall, the model captures closely lower-order correlations, the voting margin distribution, and performs well even on the difficult test of predicting the entire probability distribution of median voters.

\begin{figure*}[t!]
    \centering
    \includegraphics[width=.82\linewidth]{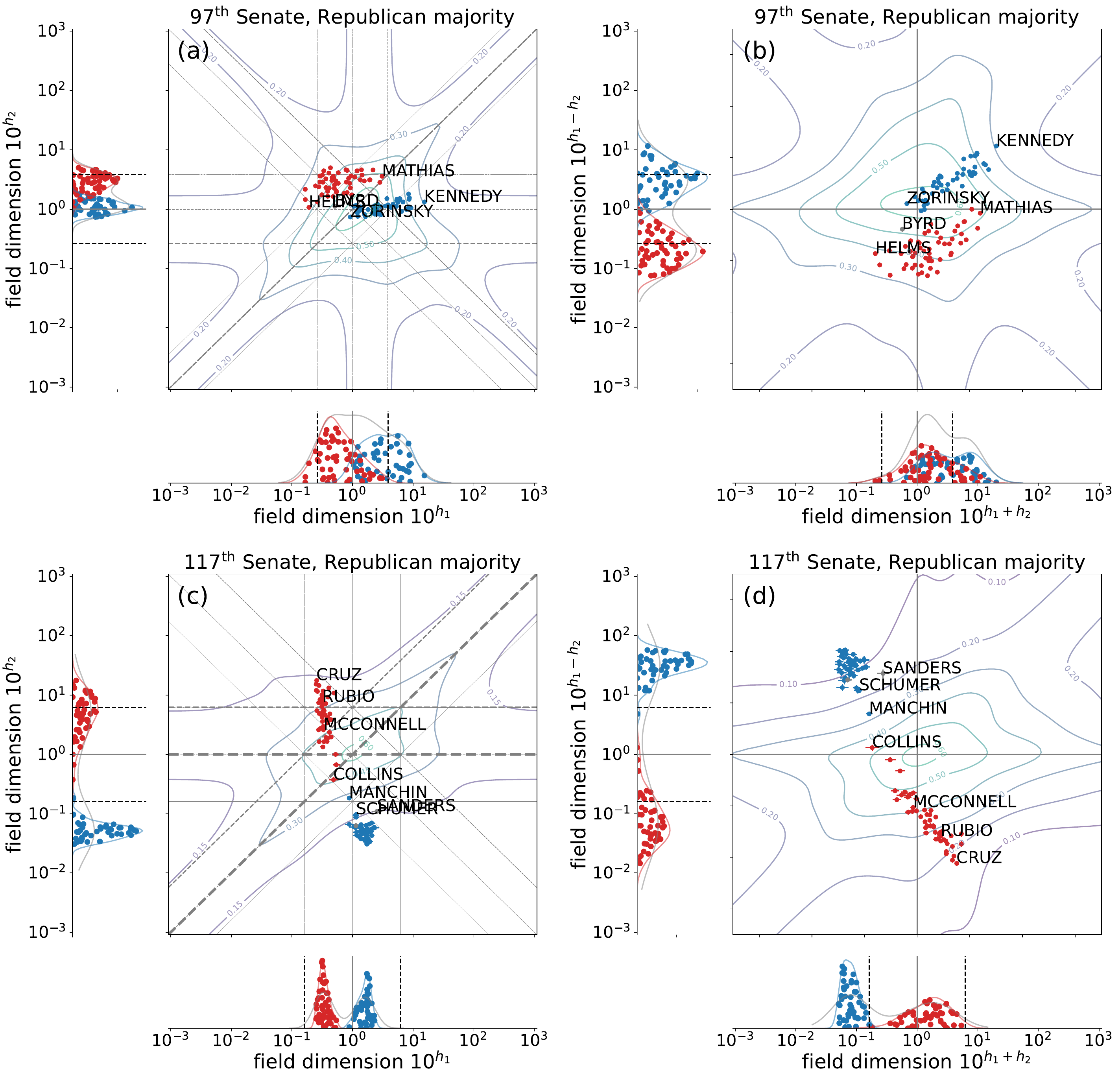}
    \caption{Maps of voter preferences for the (a, b) \session{97} and (c, d) \session{117} Senates. (a, c) Non-consensus fields and (b, d) their linear combinations. Note that the right column is the same parameters in the left column but rotated by $\pi/2$ to obtain the linear combinations. Standard error bars over posterior sample. Axes are exponentiated to show relative probabilities; i.e., a positive unit distance represents a factor 10 increase in probability of voting `yea.' Width of gray lines in panels a and c are proportional to the corresponding context probability. Density plots parallel to the coordinate axes indicate the distribution of senators along the corresponding axis. (a) Edward Kennedy (D-MA) and Charles Mathias (R-MD) in the \session{97} both have large field magnitudes, but are predisposed to bipartisanship --- consider the $h_2$ projection and $h_1$ projection, respectively. They are also proximate along $h_1-h_2$ and did collaborate \cite{rogersSensEdward1981}. (c) Sanders (I-VT) is the non-Republican most likely to vote with Republicans as is Susan Collins (R-MA) for non-Democrats along the $h_1+h_2$ projection. Ted Cruz (R-TX) is positioned at the extreme of the Republican party and is less likely to vote with bipartisan coalitions. 
    Contour lines indicate voter entropy (Eq~\ref{eq:complexity}) for voters with the given fields and other parameters sampled over the posterior distribution. Remaining Congresses shown in ``Supplementary Information.''}
    \label{fig:maps}
\end{figure*}

\section{Map of the partisan divide}
To gain insight into what the parameters mean, we visualize the minimal model that captures the essential features of the probability distribution, the $K=3$ model. We show the free fields projected onto a two-dimensional subspace, denoted $\vec\sigma_1$ and $\vec\sigma_2$, in Figure~\ref{fig:maps}. We graph the consensus field separately. First, we observe that the members of the two parties tend to cluster together, reflecting the partisan divide. The effective field along any of the two dimensions is given by the projection of the points along either the $x$- or $y$-axis. Interestingly, the projection along the $x$-axis in panel a indicates the clear partisan divide because the distribution of $h_{\rm i}^{\rm eff}$ is bimodal, but the projection along the y-axis displays overlap between the two parties, akin to a bipartisan mode that is not determined by a simple left-right split. Similar variation occurs for the $h_1+h_2$ and $h_1-h_2$ compositions of the fields that emerge when projected onto the diagonal 1:1 and 1:$-$1 lines. This observation immediately makes clear that it is possible to obtain in a multi-dimensional space the coexistence of bipartisanship and polarization, which by definition cannot occur in a unidimensional and thus zero-sum formulation \cite{blackRationaleGroup1948}.

Such a conclusion changes little when accounting for the force to consensus, $h_{\rm unan}$, the strength of which is shown by the dashed lines on either side of the axes. When comparing the \session{97} and \session{117} sessions, we see that the force to unanimity is relatively strong in the former compared to the latter, also reflecting the dominance of partisan forces in later congresses. Thus, this projected representation presents a rich but also intuitive way of considering the patterns of voting in Congress and highlights its multidimensionality.

The spread of the members of each party in the parameter space also implies that partisan coalitions cannot always satisfy all members of both parties. Imagine that the two parties were tightly confined about the point $(h_D,0)$ for the Democratic party and $(0,h_R)$ for the Republication party (besides the consensus field). Since the points are orthogonal and on an axis of symmetry, it would be possible to find contexts for which one of the two parties would be apathetic; for example, $\vec\sigma=\{1,0\}$. As a result, coalitions could ``piece'' together orthogonal issues along orthogonal axes. In this 2-dimensional example, this would be a projection along the 1:1 line, representing independent contributions to a cooperative $h_D+h_R$ outcome. The arrangement of the two parties in Figures~\ref{fig:maps}a and b reveal few simple modes like such because of how the two parties are spread over the parameter landscape in the \session{97} Congress (again because discrete rotations and reflections are the only possible ways to reorient the parties around each other). The projections along the $x$ and $y$-axes indicate that in three out of the four possible ways to combine the fields, the parties are strongly mixed. In contrast, we see a somewhat different picture in panels c and d for the \session{117} Congress. The parties pull away from each other, and the centers of mass are nearly antipodal. Three of the four density projections parallel to the coordinate axes show separated, party-aligned coalitions, rather than overlap between the red and blue points. Thus, our graphical representation of the voting statistics indicates fewer opportunities for bipartisan alignment and the predominance of antagonistic contexts, which arise not from any single individual's partisan vote, but from collective opposition (see Figures~\ref{fig:maps2}-\ref{fig:maps4}).

\section{Voter and contextual entropies}
The flexibility of individual voters and the number of different coalitions emerge as separate terms in the information entropy of the model \cite{shannonMathematicalTheory1948, coverElementsInformation2006}, one each voter and one for the context,
\begin{align}
	S[p(\vec s, \vec\sigma)] &= \sum_{\rm i=1}^N S_{\rm i} + S[P(\vec\sigma)]. \label{eq:complexity}
\end{align}
Since the voters are independent once conditioned on context $\vec\sigma$, we have factorized the first term into the contributions from each individual indexed i, where $S_{\rm i} \equiv \br{S[P(s_{\rm i}|\vec\sigma)]}_{\vec\sigma}$. Thus, the structure of the model allows us to consider each individual contribution to the total entropy $S$ separately, and only conditioned on, the context. 
To gain intuition about how voter entropy depends on voting preferences, we can calculate the entropy of an imaginary voter on the plane with the fields at the given coordinates shown in the contours of Figure~\ref{fig:maps}. As we go further out to the fringes of parameter space, where the fields become larger, voter entropy decreases as voters become more biased in their voting patterns. Near the center, voters become less predictable; thus, the entropy $S_{\rm i}$ serves as a measure of voter complexity.

\begin{figure}[t!]
    \centering
    \includegraphics[width=\linewidth]{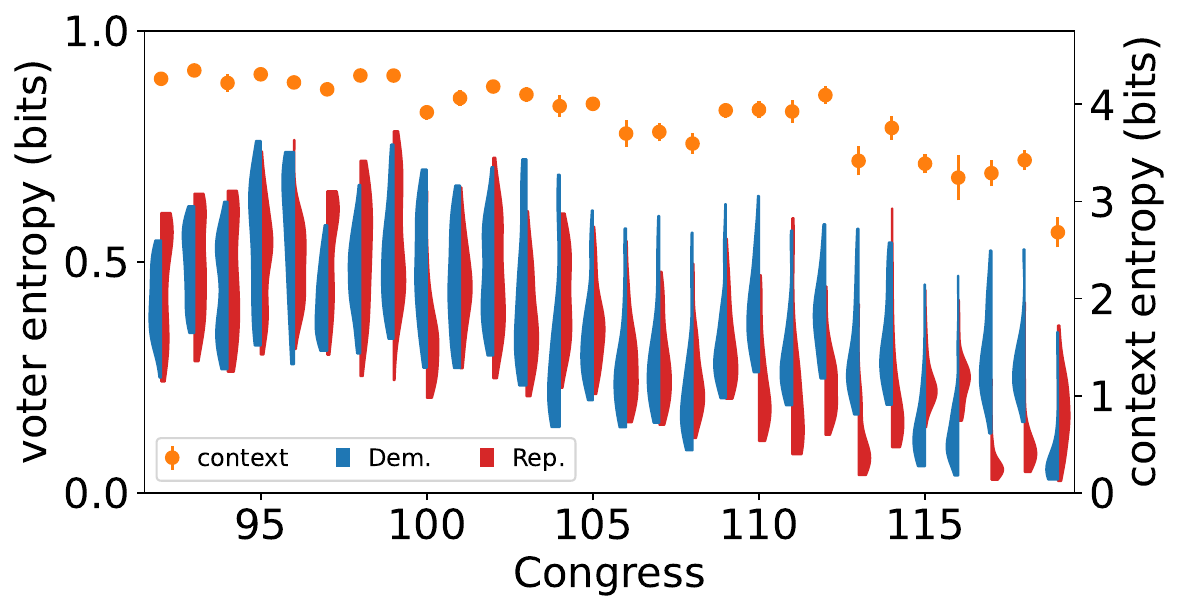}
    \caption{Voter entropy $S_{\rm i}$ over time for $K=3$ model. Violin plots show distributions of $S_{\rm i}$ over Democratic (blue) and Republican (red) senators. They trend lower and the distributions separate from each other in recent congresses. An early example is the \session{100} Senate, the last years of George W.~Bush's presidency, during which both houses of Congress were controlled by the Democrats. Context entropy shown with orange markers.}
    \label{fig:complexity}
\end{figure}

The isocontours indicate a change in effective fields that redistributes yea and nay votes in a way that maintains voter complexity. In contrast, paths that cross isocontours would change their balance of yea and nay votes, and this would preferentially favor one party at the expense of the other (Figure~\ref{gr:vote margin}). Thus, cases of short paths that lie parallel to isocontours indicate an interesting symmetry: a voter could be replaced by a hypothetically similar individual that appears in the other party, but remain indifferent in the number of times they are agreeing or disagreeing with bills overall. This kind of flexibility is possible to some extent in the \session{97} Congress (see Mathias as an example). In the \session{117} Congress, such a shift must take a long detour because isocontours separate the parties, presenting a topographical depiction of the partisan divide, where the Democrats are ``well-defended'' by a ridge of voting complexity. This presents a stronger form of the partisan separation that we find in the \session{97} session because it indicates the emergence of a divide that, in a closely divided Senate, can only be directly crossed by changing the balance of yea and nay and preferentially favoring one party at the expense of the other. 

In addition, the voter complexity contours reveal intraparty diversity because they show how relevant the spread in parameter space is to overall voting record. The closer the voters are to the peaks in entropy, the more unexpected their voting records are from the perspective of a simple partisan description. Indeed, we find swing voters like Collins (R-ME) and Manchin (D-WV) closer to the peaks and ``hard partisans'' like Kennedy (D-MA), East (R-NC), Cruz (R-TX), and Paul (R-KY) on the ``flat plains.'' In the \session{117} Senate, the vast body of the Democratic majority lies on the plains with very low voter complexity, so while it is the case that they may have contrasting effective fields, they have similar average votes (Figure~\ref{gr:vote margin}); this indicates the cohesion amongst the Democratic minority. Looking more closely, we see that Manchin (D-WV) is situated between the two parties along every projection; he occupies a special median position, always located in between the two parties but not in any way strongly aligned with the core of either. In contrast, Sanders (I-VT), shows dichotomy that is not indicative of a classic swing voter. Along one projection, he is on the far wing of the Democrats ($h_1-h_2$), even more than Collins, and along another he is the Democrat most likely to vote with Republicans ($h_1+h_2$). This indicates an unusual voting record in this session, and correspondingly his entropy is amongst the highest in the Democratic party. Thus, these maps provide an intuitive but empirical and multidimensional portrait of voting that may be surprising in light of partisan projections that dominate popular discourse \cite{fischerThingsAre2024} (and often the literature).

As the widening spread of points on the maps suggest, voter complexity generally decreases over time as would be consistent with an account of increasing polarization. Previous to the \session{105} session, voter complexity is comparable across the two parties and likewise issue complexity is high as we show in Figure~\ref{fig:complexity}. Afterwards, voter complexity separates by majority-minority party, voter complexity overall drops, and issue complexity trends down. The pattern is quite striking especially in contrast to Figure~\ref{gr:vote margin} from which we obtain little indication of partisan polarization until the \session{113} Congress. 
Furthermore, the drop in issue entropy of about 1 bit for $K=3$ (and 1.5 bits for $K=4$) dominates the drop in voter entropy of about 0.3 bits (or 0.4 bits for $K=4$). This suggests that polarization is better explained in our model by the choice of in contexts in which the votes occur. Furthermore, the context entropy remains in the range of several bits for $K=3$ and $K=4$, which correspond contextual configurations numbering on the order of 10. While this seems limited in the face of the wider space of policy issues, it is far from a unidimensional, zero-bit scenario.

\begin{table}
\centering
\caption{Equivalence classes between conditionally independent voter and interaction models. Mapping is obtained by marginalizing out the context $\vec\sigma$ and its distribution $P(\vec\sigma)$, for either discrete or continuous spins $\vec s$. The reduction to Ising-type, pairwise interactions then holds at leading order in a weak-field approximation (see Supplementary Information). The circle and spherical surface entries are expansions in the corresponding Bessel functions, whose coefficients are fixed by the anisotropy of the fields and by the orientation of the total field relative to its principal axes. The uniform distribution over discrete $\vec\sigma$ would be a rudimentary version of the models discussed in previous work and this work. We detail each case in ``Supplementary Information.''}\label{tab:classification}
\begin{tabular}{c||c|c|}
	\multicolumn{1}{c}{} & \multicolumn{2}{c}{\textbf{voters or spins $\vec s$}} \\
	\parbox[c]{2.2cm}{\centering\textbf{distribution $P(\vec\sigma)$\\ and context $\vec\sigma$}\vspace{.1cm}} & discrete & continuous\\
\toprule
	Gaussian and real & Ising, Potts & XY, Heisenberg \\
\hline
	uniform and discrete & \parbox[c]{2.2cm}{\centering \vspace{.1cm}Lee-Cantwell \cite{leeValenceInteractions2024}, \\ Lee\vspace{.1cm}} & unexplored \\
\hline
	uniform over circle & \multicolumn{2}{c|}{modified Bessel expansion} \\
\hline
	uniform over sphere & \multicolumn{2}{c|}{modified spherical Bessel expansion} \\
\hline
\end{tabular}
\end{table}

\section{Equivalence to interactions}
Although we started by contrasting interaction from voter preference models, there is a deep connection between the two formulations \cite{kruisThreeRepresentations2016, bulsoRestrictedBoltzmann2021a}. Consider a simple example of the single-issue model with context $\sigma\in\{-1,1\}$. When the two orientations occur with equal probability,
\begin{align}
\begin{aligned}
P(\vec s) &= \frac{1}{2Z}\left( e^{\sum_{\rm i} h_{\rm i}s_{\rm i}} + e^{-\sum_{\rm i} h_{\rm i}s_{\rm i}}\right)\\
	 &= \frac{1}{Z'} \exp\left[\sum_{\rm i<j} \tilde J_{\rm ij} s_{\rm i}s_{\rm j} + \sum_{\rm i<j<k<l}\tilde L_{\rm ijkl} s_{\rm i}s_{\rm j}s_{\rm k}s_{\rm l} + \cdots\right],
\end{aligned}\label{eq:equivalence relation}
\end{align}
incurring a set of higher-order interaction terms that correspond to effective pairwise, quartet, and remaining even-order terms in the Hamiltonian.\footnote{In \eqref{eq:equivalence relation}, for example, we can express $\tilde J_{\rm ij}$ as a series expansion in powers of $t_{\rm i}\equiv\tanh h_{\rm i}$, $\tilde J_{\rm ij} = t_{\rm i}t_{\rm j}\left(1 - \sum_{\rm k\neq i,j} t_{\rm k}^2\right) + O(t^6)$.} As a result of the transformation, $Z$ and $Z'$ differ by a multiplicative constant. When the probability of the two orientations of $\sigma$ are not equally 1/2, we would obtain a mixture model now with interactions of odd-order (see ``Supplementary Information''). This is a simple RBM with one, multi-dimensional hidden unit $\vec\sigma$.

More generally, the symmetries of the contextual space $\vec \sigma$ determine the interaction structure \cite{sejnowskiLearningSymmetry1986}, besides the cubic lattice that we consider here (this is in addition to the invariances encapsulated in Eq~\ref{eq:symmetries}). If $\vec\sigma$ displays rotational symmetry and the distribution of $Q(\vec\sigma)$ is multi-variate Gaussian, then in the small $h_{\rm i}$ limit we obtain the Hopfield model, which is used for describing neural statistics \cite{hopfieldNeuralNetworks1982}. If the spins are continuous with Gaussian $Q(\vec\sigma)$, we recover the XY or Heisenberg models. With Eq~\ref{eq:K model} and uniform distribution $Q(\vec\sigma)$, we obtain a linear combination of of nonlinear basis functions, the hyperbolic cosines, which as we describe above corresponds to an even-order interaction expansion. Under uniform $P(\vec\sigma)$ and restricting $\vec\sigma$ to a coordinate on the unit circle, we would recover a sum over modified Bessel functions, an analog to the XY model as we summarize in Table~\ref{tab:classification}. On a sphere, we again obtain Bessel functions (see ``Supplementary Information'' Appendix~\ref{si sec:interactions} for more details). Thus, interactions between spin models in statistical mechanics are a result of the symmetries assumed in the hidden units as we enumerate in  Table~\ref{tab:classification}, a classic idea that we generalize to a nontrivial hidden context $Q(\vec\sigma)$ to present a powerful technique for capturing other higher-order correlations with minimal parameterization.

\section{Discussion}
In statistical physics, we often think of collective order as emerging from local interactions. In contrast, order is often assumed into the individual in models of political voting through features like ideological preference, individual strategy, or utility. What we show here how both pictures can be consistent with one another with the restricted Boltzmann machine (RBM). 
Building on the long history of foundational work on voting and opinion dynamics in statistical physics \cite{krapivskyDynamicsMajority2003, galamRoleInflexible2007, castellanoStatisticalPhysics2009, baronchelliVoterModels2011, xieSocialConsensus2011, galesicStatisticalPhysics2019, rednerRealityinspiredVoter2019, phamEmpiricalSocial2022}, we infer heterogenous model parameters directly from data.
The core idea behind the model is that each voter has a vector of preferences along $K$ different dimensions, which correspond to shared aspects of the bill, which we call ``issues'' to evoke the picture of building coalitions by adding elements to a bill or by modifying voting rules. More generally, this refers to context of the vote shared by all the voters. Then, preferences can constructively add up or interfere destructively to push the voter to vote either `yea' or `nay.' As a result, the models incorporate individual preferences (a key tenet of political voting theory), are minimal, and yet are expressive enough to capture higher-order correlations.

Moving to the language of RBMs, the context is the hidden unit and the vote is the visible unit. The weights connecting the hidden and visible units are vectors, not scalars as usual, consisting of the $K$-dimensional preferences. It is the dot product between the context and preference vectors that endows each voter with a vote tendency towards `yea' or `nay.' 
When applied to roll-call voting in the U.S.~Senate, we find that weights with dimensionality ${K=3}$ closely capture nontrivial statistics like the distributions of winning margins and the statistics of median voters with small improvements to performance when moving to $K=4$ (Figures~\ref{gr:vote margin}, \ref{fig:observables}, and \ref{gr:fit}). These values of $K$ correspond to 3-4 bits or 4-5 bits, respectively, for the entropy of the hidden units, representing the variability of context, even in the latest Congresses. This indicates that the order of 10 contexts are sufficient --- and substantially more than one required as in the unidimensional partisan case to support the classic ideological zero-sum game \cite{blackRationaleGroup1948} --- to model senatorial voting. 

The parameters transparently relate to the probabilities of voting, meaning that they provide an unfiltered compression of the voting record. To leverage this, we devise a two-dimensional visualization for the $K=3$ model. Figure~\ref{fig:maps} highlights the manner in which notable voters sit between the two parties. Senator Bernie Sanders, for example, is the least likely to vote with Republicans in one context and yet the most likely to vote with the Republican party in another. In contrast, Senator Joe Manchin is well-positioned between  the two parties as a median across all contexts. While it remains to relate the context vectors $\vec\sigma$ to particular elements like line items in a bill or the rules of the vote, the relative difference in the senators' locations, and especially how Sanders swaps from one side to another, signals alternative strategies for spanning the political divide. This could not be distinguished in a single dimension \cite{giansiracusaComputationalGeometry2019}.

The map also depicts ways in which the partisan divide has become stronger. Voters move out from the center as strengthening fields indicate increasing voter inflexibility. As they do so, they show reduced voter complexity, which we define as a measure of their imbalance of `yea' or `nay' votes. As voters move away from the center, the contours of voter complexity also shift to reveal a topographical barrier between the two parties. The two parties are more often separated by a ridge of high entropy in recent sessions, which reduces voter parity: fewer pairs of voters across the partisan aisle can be considered similar. Finally, the way that the parties are positioned relative to one another in the space indicates that there are fewer contexts for bipartisan votes. It is both these elements, individual and collective contributions, that emerge as signatures of change from the voter preference maps.

The individual and collective aspects naturally segregate in the information entropy, allowing us to calculate their respective contributions to polarization. While both decrease over time in Figure~\ref{fig:complexity}, the larger drop occurs in the context entropy. This suggests an important and, perhaps in popular discourse, underappreciated aspect of voting, which is the impact of what politicians choose to vote on and the procedures used to determine them matter such as the ``rule of four'' for deciding when to hear an appeal in the U.S.~Supreme Court or the way that the voting agenda is set in Congress \cite{theriaultPartyPolarization2008}. 

We draw and extend upon foundational models in statistical physics, originally motivated by emergent collective computational properties of neural networks \cite{hopfieldNeuralNetworks1982, hintonBoltzmannMachines1984}. We demonstrate an example of how they present an avenue for rich extension through the geometry of the hidden units on the example of the U.S.~Senate --- presenting an opening example to a much wider list of example systems that have been explored by the maxent framework, ranging from neural networks to collective motion \cite{schneidmanWeakPairwise2006, moraMaximumEntropy2010, moraMaximumEntropy2010, coccoAdaptiveCluster2011, bialekStatisticalMechanics2012, shemeshHighorderSocial2013, schwabZipfsLaw2014, leeStatisticalMechanics2015, merchanSufficiencyPairwise2016}. Essentially, the assumed geometry corresponds to a lower-dimensional representation of interactions \cite{coccoHighdimensionalInference2011} (Table~\ref{tab:classification}), a mirror to the search for hidden symmetries \cite{sejnowskiLearningSymmetry1986}. In this sense, the equivalence is a mathematical curiosity, or even an opening for learning model parameters more efficiently. On the other hand, the debate about the physical meaning of such interactions takes on greater implications for models of social systems, where the question of what factors could be responsible for qualitative changes in social dynamics has larger implications \cite{danielsControlFinite2017}. Indeed, it seems important to consider not only interventions at the level of the individual \cite{mobiliaDoesSingle2003, campbellLargeElectorates1999, iacopiniGroupInteractions2022, xieSocialConsensus2011, axelrodPreventingExtreme2021}, where opinions often appear to change weakly \cite{zwickerPersistentBeliefs2020, stanleyResistancePosition2020}, and to consider the impact of changes at the collective level \cite{fiorinaUnstableMajorities2017}.

\matmethods{
\noindent\textbf{Data}

We combine data from the VoteView project and the congress-legislators project on Github (\url{https://github.com/unitedstates/congress-legislators}). We only keep the set of senators who have participated in the maximum number of roll-call votes of any other, which removes transient Senators including those who served temporarily due to special elections. This retains about $94\%$ of all roll-call votes cast and 97-100 senators in the congressional session of interest except in the \session{104}, \session{111}, and \session{115} with 96, 91, and 96 senators, respectively. While we focus on roll-call votes because they tally votes by individual, they suffer from selection biases and procedural rules which can complicate the interpretation of some behaviors, limitations which are well noted in the literature \cite{robertsStatisticalAnalysis2007}. We do not explicitly consider the voters' absences in our model, however, absences can occur with a non-negligible probability in our dataset: the average Senator will be absent for 5-10\% of Senate roll-call votes. When testing the distribution of winning margin $p(k)$, we use exactly the pattern of absences that are seen in the data to insert absences into the posterior sample, or the assumption that the pattern of absences is independent of the votes.

\vspace{.1cm}
\noindent\textbf{Sampling from the posterior}

We sample from the posterior using the Hamiltonian Monte Carlo No U-Turn Sampler (HMC NUTS) implemented in NumPyro. We run $10^3$ independent MCMC chains initialized with random samples from the prior and run them for $T=5\times10^3\times K$ steps with tree search depth of 6. Code will be provided in a GitHub repository to reproduce our sampling procedure.

We find from numerical exploration that an unconstrained fields problem generally performs worse in terms of likelihood and lower-order statistics like the average votes and pairwise correlations. 
The large parameter space is difficult to search exhaustively; thus, the consensus field is \textit{a priori} unnecessary, but it substantially reduces the parameter space, improves the quality of the solutions found by reducing the space of possible solutions, and enforces a kind of minimality that is also consistent with a shared, bipartisan tendency to vote in the same way. While removing symmetries and additional simplifying assumptions do not necessarily remove the problem of multiple maxima in the solution landscape, we find from extensive numerical sampling that for many of the congressional sessions only a single peak in the posterior emerges, which demonstrates that the approach leads to a better posed problem than fully free biases. This is reflected in the sampled parameters, whose standard error of the mean converges to a stable value as we show in Figure~\ref{fig:maps}. 

\vspace{.1cm}
\noindent\textbf{Collapsing degeneracies in the posterior}

An obstacle to sampling from the posterior is that the symmetries of $\vec\sigma$ identify degeneracies in the solution landscape. As an example, if we reverse the sign of all the fields $\vec h_{\rm i}\rightarrow -\vec h_{\rm i}$ and we reverse the sign of all the issue vectors $\vec\sigma\rightarrow-\vec\sigma$, then we have preserved the value of the Hamiltonian because the effective fields $h_{\rm i}^{\rm eff}$ remain unchanged. More generally, any transformation of the fields that preserves the set of effective fields up to a permutation leads to a degenerate solution. 

To formulate the general problem, we write the matrix of all effective fields $\mathcal H$ in each possible configuration of $\vec\sigma$, represented as the matrix $\Sigma \equiv \{\vec\sigma\}$, 
\begin{align}
	\mathcal H &= \Sigma H. \label{eq:eff H}
\end{align}
Eq~\ref{eq:eff H} defines the collection of effective fields that result from projecting the fields $H$ onto all configurations in $\Sigma$. Now, we seek an operator $\fancyR$ that transforms the fields $H$ while preserving the effective fields $\mathcal H$. Then, the condition of invariance stipulates that the result $\Sigma \mathcal{R} H$ must only permute the rows of $\mathcal H$ (otherwise the probability distribution $P(\vec s)$ will not in general remain invariant), $\fancyP \mathcal H = \Sigma\,\fancyR H$.
For permutations $\fancyP^{-1} = \fancyP^T$. Then,
\begin{align}
	\mathcal H &= \fancyP^T \Sigma\,\fancyR H. \label{eq:symmetries}
\end{align}
Thus, operators $\fancyR$ are permitted that correspond to an inverse permutation of the rows of $\Sigma$. The only such allowed operations are discrete rotation symmetries (multiples of $\pi/2$) and reflections. Note that these are unitary operations that would preserve the rank of $H$ and cannot eliminate or add to any of its symmetries.

To compare independent MCMC chains from the posterior distribution, we must collapse the degeneracy. For the jth sample from the posterior distribution, we enumerate all symmetry transformations to find the most similar transformation $(\fancyP, \fancyR)$ to minimize the summed KL divergence between two chains $\rm k=0$ and $\rm k>0$ for the individual voter probability distributions $P(s_{\rm i})$, notated $D_{\rm i}$, and the pairwise probability $P(s_{\rm i}, s_{\rm j})$, notated $D_{\rm ij}$. The total loss function we minimize is then $\sum_{\rm i=1}^N D_{\rm i} + \sum_{\rm i<j}^N D_{\rm ij}$. This presents a heuristic to account for the degeneracies originating from symmetry transformations in the solution landscape.

\vspace{.1cm}
\noindent\textbf{Spatial voting}

We might take a moment to consider how our physics-inspired approach is related to the dominant and highly influential model in the political science literature, W-NOMINATE \cite{pooleScalingRoll2011, pooleSpatialModel1985, lewisVoteviewCongressional2025}, and a model that is almost exclusively cited in popular discourse about polarization in political voting. At a superficial level, our approach is analogous because they also assume that individuals have parameterized voting tendencies that are expressed across a finite number of bill dimensions. In contrast, W-NOMINATE assumes a form of a utility function \cite{downsEconomicTheory1957, pooleSpatialModel1985}, and, during the model fitting process, seeks the locations of the voters in a high-dimensional space and the corresponding locations for each bill that maximizes the fit to the data. For each bill, one must find the parameters that describe it and the corresponding weights of each individual along each dimension. However, unanimous votes are excluded from the model (despite their importance in the distribution), the number of parameters dwarfs the number of data points for similar accuracy, and the complexity of the model means that it is difficult to understand how the parameters relate to voting patterns. The approach we present here is parsimonious and explicitly accounts for model complexity while presenting a clear, tractable, and accurate approach to modeling voter preferences.
}
\showmatmethods{}

\acknow{
We thank a research intern for pilot work and computational efforts and George Cantwell, Peter Steiglechner, and Gabriela Juncosa for helpful discussion and feedback. This research was funded in whole or in part by the Austrian Science Fund (FWF) 10.55776/ESP127. We acknowledge funding from the International Communication program of the Austrian Research Association (OEFG). This work was co-financed by the Austrian Federal Ministry for Innovation, Mobility and Infrastructure (BMIMI) as part of the project GZ 2021-0.664.668. This work used AI.
}
\showacknow{}

\bibliography{refs} 

\clearpage

\appendix

%
%

\section{Derivation of interactions for some context geometries}
\label{si sec:interactions}

Case of continuous one-dimensional $\sigma$ with Gaussian distribution.
\begin{align}
P(\vec s) &= \int_{-\infty}^\infty d\sigma\, p(\vec s|\sigma) p(\sigma)\\
 &= \int_{-\infty}^\infty d\sigma \frac{e^{\sigma\sum_{\rm i} h_{\rm i}s_{\rm i}}}{2^N \prod_{\rm i} \cosh(\sigma h_{\rm i})} \frac{e^{-\sigma^2}}{\sqrt{\pi}}\\
 &= \frac{1}{2^N \sqrt{\pi}}\int_{-\infty}^\infty d\sigma\, e^{-\sigma^2 - \sum_{\rm i}\log \cosh(\sigma h_{\rm i}) + \sigma\sum_{\rm i} h_{\rm i}s_{\rm i}}
 \end{align}
 From the power series expansion of $\log \cosh$, when $h_{\rm i}$ are small, we can approximate $\log \cosh (\sigma h_{\rm i}) \approx \frac{(\sigma h_{\rm i})^2}{2}$, so that with $a \equiv 1 + \frac{1}{2}\sum_{\rm i} h_{\rm i}^2$
 \begin{align}
P(\vec s) &\approx \frac{1}{2^N \sqrt{\pi}}\int_{-\infty}^\infty d\sigma\, e^{-a\sigma^2 + \sigma\sum_{\rm i} h_{\rm i}s_{\rm i}} \\
 &= \frac{1}{2^N \sqrt{a}}\, e^{\left(\sum_{\rm i} h_{\rm i} s_{\rm i}\right)^2\big / 4a}\\
&= \frac{1}{2^N \sqrt{a}}\, e^{\sum_{\rm i} h_{\rm i}^2 / 4a}\, e^{\sum_{\rm i<j}J_{\rm ij}s_{\rm i}s_{\rm j}}.
\end{align}
The effective interactions are $J_{\rm ij} = h_{\rm i}h_{\rm j}/2a$.

The same calculation carries over to a $K$-dimensional context, which is the case relevant to Eq~\ref{eq:K model}. Let $\vec\sigma \in \mathbb{R}^K$ carry the weight $e^{-|\vec\sigma|^2}$, give each voter a field vector $\vec h_{\rm i}$, and collect the fields into $\vec H \equiv \sum_{\rm i} s_{\rm i}\vec h_{\rm i}$, so that
\begin{align}
	P(\vec s) &= \frac{1}{2^N Z}\int d^K\sigma\, e^{-|\vec\sigma|^2 - \sum_{\rm i}\log\cosh\left( \vec h_{\rm i}\cdot\vec\sigma \right) + \vec H\cdot\vec\sigma}
\end{align}
The same weak-field approximation replaces $\log\cosh(\vec h_{\rm i}\cdot\vec\sigma)$ by $(\vec h_{\rm i}\cdot\vec\sigma)^2/2$, and since $\sum_{\rm i}(\vec h_{\rm i}\cdot\vec\sigma)^2 = \vec\sigma^{\mathsf T}M\vec\sigma$ for the field second-moment matrix
\begin{align}
	M &\equiv \sum_{\rm i=1}^N \vec h_{\rm i}\vec h_{\rm i}^{\mathsf T}
\end{align}
the scalar $a$ of the one-dimensional case becomes a matrix $A \equiv I + \frac{1}{2}M$. Performing the Gaussian integral and using $s_{\rm i}^2=1$ to separate the diagonal terms as before,
\begin{align}
	P(\vec s) &= \frac{\pi^{K/2}}{2^N Z \sqrt{\det A}}\, e^{\vec H^{\mathsf T}A^{-1}\vec H/4}\\
	&= \frac{\pi^{K/2}\, e^{\frac{1}{4}\sum_{\rm i}\vec h_{\rm i}^{\mathsf T}A^{-1}\vec h_{\rm i}}}{2^N Z \sqrt{\det A}}\, e^{\sum_{\rm i<j}J_{\rm ij}s_{\rm i}s_{\rm j}}
\end{align}
with the coupling
\begin{align}
	J_{\rm ij} &= \frac{1}{2}\vec h_{\rm i}^{\mathsf T}\left( I + \tfrac{1}{2}M \right)^{-1}\vec h_{\rm j} \label{eq:Jmulti}
\end{align}
which reduces to $J_{\rm ij}=h_{\rm i}h_{\rm j}/2a$ when $K=1$.

Eq~\ref{eq:Jmulti} makes the connection to the Hopfield model precise. By definition, Hopfield couplings are Hebbian, $J_{\rm ij}\propto\vec h_{\rm i}\cdot\vec h_{\rm j}$, and Eq~\ref{eq:Jmulti} takes that form only when $A^{-1}$ is proportional to the identity, that is only when $M$ is isotropic. When $M$ is anisotropic the directions in context space carrying the most field are screened the most strongly, and the coupling is a distorted Hebbian form rather than a rescaled one. In the strict limit of vanishing fields $M\to0$ and $A\to I$, so the Hebbian form is always recovered in the weak-field approximation.

More carefully, the approximation is valid when the discarded quartic term of $\log\cosh$, namely $\frac{1}{12}\sum_{\rm i}(\vec h_{\rm i}\cdot\vec\sigma)^4$ is much smaller than the quadratic term. Under the Gaussian measure that survives, $\vec\sigma$ has covariance $A^{-1}/2$, so with
\begin{align}
	g_{\rm i} &\equiv \vec h_{\rm i}^{\mathsf T}A^{-1}\vec h_{\rm i}
\end{align}
we have $\langle(\vec h_{\rm i}\cdot\vec\sigma)^2\rangle = g_{\rm i}/2$ and, by Wick's theorem, $\langle(\vec h_{\rm i}\cdot\vec\sigma)^4\rangle = 3g_{\rm i}^2/4$. The ratio of the quartic to the quadratic contribution from voter i is then
\begin{align}
	\frac{\tfrac{1}{12}\langle(\vec h_{\rm i}\cdot\vec\sigma)^4\rangle}{\tfrac{1}{2}\langle(\vec h_{\rm i}\cdot\vec\sigma)^2\rangle} &= \frac{g_{\rm i}}{4}
\end{align}
so that the expansion is controlled by $\max_{\rm i} g_{\rm i}\ll4$ rather than by the total field.

The distinction matters because $g_{\rm i}$ is bounded. Using $A^{-1}M = 2\left( I - A^{-1} \right)$,
\begin{align}
	\sum_{\rm i=1}^N g_{\rm i} &= \textrm{tr}\left( A^{-1}M \right) = 2\left( K - \textrm{tr}\,A^{-1} \right) < 2K \label{eq:gbudget}
\end{align}
The context offers only $K$ directions in which field can be absorbed, and that fixed budget is shared among all $N$ voters. What enters the quartic, however, is $\sum_{\rm i}g_{\rm i}^2$ and not $\sum_{\rm i}g_{\rm i}$, and the same budget may be spread thinly over many voters or concentrated on a few. Defining a ``participation ratio'', we have
\begin{align}
	N_{\rm eff} &\equiv \frac{\left( \sum_{\rm i}g_{\rm i} \right)^2}{\sum_{\rm i}g_{\rm i}^2}\label{eq:participation ratio}
\end{align}
\eqref{eq:participation ratio} equals $N$ when every voter carries field equally and $1$ when a single voter carries all of it, lying between those extremes by the Cauchy--Schwarz inequality. Since $A$ is positive definite, Eq~\ref{eq:gbudget} may be squared to give $\left( \sum_{\rm i}g_{\rm i} \right)^2<4K^2$. Rearranging the definition of $N_{\rm eff}$ then bounds the aggregate weight of the discarded quartic,
\begin{align}
	\sum_{\rm i}g_{\rm i}^2 &= \frac{\left( \sum_{\rm i}g_{\rm i} \right)^2}{N_{\rm eff}} < \frac{4K^2}{N_{\rm eff}}
\end{align}
For $K=1$ this reads $\sum_{\rm i}h_{\rm i}^4\big/a^2 < 4/N_{\rm eff}$. The collapse to pairwise interactions is therefore controlled by the number of voters rather than by the strength of the fields, and it improves as voters are added at fixed field per voter. For $N$ of the order of the size of the Senate it is accurate even when $\sum_{\rm i}h_{\rm i}^2$ is not small, which is the regime the fitted fields occupy.

Consider the circumference of the circle, assuming radius $r=1$, with $\vec\sigma$ uniform in $\theta$ and each voter carrying a field vector $\vec h_{\rm i}$. As in the Gaussian case, we have
\begin{align}
	P(\vec s) &= \frac{1}{2^N}\, \frac{1}{2\pi}\int_0^{2\pi} d\theta\; e^{-\sum_{\rm i}\log\cosh\left( \vec h_{\rm i}\cdot\vec\sigma \right) + \vec H\cdot\vec\sigma}
\end{align}
with $\vec H = \sum_{\rm i}s_{\rm i}\vec h_{\rm i}$ as above. No separate normalization is required, since the conditional and the uniform measure are each normalized already.

The same weak-field approximation gives $\sum_{\rm i}\log\cosh(\vec h_{\rm i}\cdot\vec\sigma)\approx\frac{1}{2}\vec\sigma^{\mathsf T}M\vec\sigma$ with the same second-moment matrix $M$, and on the unit circle that quadratic form separates into a constant and a single harmonic,
\begin{align}
	\vec\sigma^{\mathsf T}M\vec\sigma &= \frac{\textrm{tr}\,M}{2} + \frac{\Delta}{2}\cos\left( 2\left( \theta - \theta_M \right) \right)
\end{align}
where $\Delta \equiv \lambda_1 - \lambda_2$ is the gap between the eigenvalues of $M$ and $\theta_M$ the direction of its leading eigenvector. The constant factors out of the integral; the harmonic does not, and it is what breaks the rotational invariance of the integrand.

If $M$ is isotropic then $\Delta=0$, the harmonic is absent, and writing $\vec H\cdot\vec\sigma = |\vec H|\cos(\theta-\theta_H)$ leaves
\begin{align}
	P(\vec s) &= \frac{e^{-\textrm{tr}M/4}}{2^N}\, I_0\left( |\vec H| \right)
\end{align}
for modified Bessel functions of the first kind $I_\alpha$ of order $\alpha=0$. If $M$ is anisotropic the result depends on the direction of $\vec H$ and not on its length alone. Expanding both exponentials with the Jacobi--Anger identity and using orthogonality of the harmonics,
\begin{align}
	P(\vec s) &= \frac{e^{-\textrm{tr}M/4}}{2^N}\left[ I_0\left( |\vec H| \right) I_0\!\left( \tfrac{\Delta}{4} \right) + \right.\nonumber\\
	&\left. 2\sum_{m=1}^\infty (-1)^m I_{2m}\left( |\vec H| \right) I_m\!\left( \tfrac{\Delta}{4} \right)\cos\left( 2m\left( \theta_H - \theta_M \right) \right) \right],
\end{align}
which returns to the previous expression when $\Delta=0$. The corrections are set by $\Delta$, so isotropy of $M$ is once more the condition under which the simple Bessel form survives.

Consider the spherical surface $\mathcal S$ with $\vec\sigma$ uniform on it. The same construction gives
\begin{align}
	P(\vec s) &= \frac{1}{2^N}\, \frac{1}{4\pi}\int_{\mathcal S} d\mathcal S\; e^{-\sum_{\rm i}\log\cosh\left( \vec h_{\rm i}\cdot\vec\sigma \right) + \vec H\cdot\vec\sigma}
\end{align}
and under the weak-field approximation $|\vec\sigma|=1$ lets the quadratic form be split into a constant and a traceless remainder,
\begin{align}
	\vec\sigma^{\mathsf T}M\vec\sigma &= \frac{\textrm{tr}\,M}{3} + \vec\sigma^{\mathsf T}\tilde M\vec\sigma, & \tilde M &\equiv M - \frac{\textrm{tr}\,M}{3}I
\end{align}
If $M$ is isotropic then $\tilde M = 0$, the integrand depends on $\vec\sigma$ only through $\vec H\cdot\vec\sigma$, and aligning the polar axis with $\vec H$ reduces the surface integral to an elementary one. Substituting $u=\cos\phi$,
\begin{align}
	P(\vec s) &= \frac{e^{-\textrm{tr}M/6}}{2^N}\, \frac{1}{2}\int_{-1}^{1} du\; e^{|\vec H|u} = \frac{e^{-\textrm{tr}M/6}}{2^N}\, \frac{\sinh|\vec H|}{|\vec H|},
\end{align}
with $\sinh x/x$, the modified spherical Bessel function $i_0(x)$ of order zero.

If $M$ is anisotropic the traceless part survives and the result acquires a dependence on the orientation of $\vec H$ relative to the principal axes of $M$. The natural basis is now spherical rather than the cylindrical one used for the circle. Writing the exponential in the modified spherical wave expansion,
\begin{align}
	e^{\vec H\cdot\vec\sigma} &= \sum_{\ell=0}^\infty \left( 2\ell+1 \right) i_\ell\left( |\vec H| \right) P_\ell\left( \hat H\cdot\vec\sigma \right)
\end{align}
for Legendre polynomials $P_\ell$, modified spherical Bessel functions $i_\ell$, and $\hat H = \vec H/|\vec H|$, and defining the projections
\begin{align}
	a_\ell\left( \tilde M, \hat H \right) &\equiv \frac{1}{4\pi}\int_{\mathcal S} d\mathcal S\;
	e^{-\frac{1}{2}\vec\sigma^{\mathsf T}\tilde M\vec\sigma}\, P_\ell\left( \hat H\cdot\vec\sigma \right)
	\label{eq:alsphere}
\end{align}
in direct analogy with the $c_n$ of the circle, the surface average may be taken term by term to give
\begin{align}
	P(\vec s) &= \frac{e^{-\textrm{tr}M/6}}{2^N} \sum_{\ell=0}^\infty \left( 2\ell+1 \right)
	i_\ell\left( |\vec H| \right) a_\ell,
	\label{eq:sphereexp}
\end{align}
which is exact. Under $\vec\sigma\to-\vec\sigma$ the quadratic form is unchanged while $P_\ell\to(-1)^\ell P_\ell$, so $a_\ell$ vanishes for every odd $\ell$ and only even orders contribute, exactly as only even harmonics contributed on the circle. When $M$ is isotropic, $\tilde M=0$ and $a_\ell=\delta_{\ell0}$, so the sum collapses to the single term $i_0(|\vec H|)$ obtained above.

For weak anisotropy the leading correction is explicit. Because $\tilde M$ is symmetric and traceless, $\vec\sigma^{\mathsf T}\tilde M\vec\sigma$ is a pure $\ell=2$ function on the sphere. Expanding Eq~\ref{eq:alsphere} to first order accordingly leaves $a_0=1$, $a_2 = -\frac{1}{10}\hat H^{\mathsf T}\tilde M\hat H$, and $a_\ell = O(\tilde M^2)$ for $\ell\geq4$, so that Eq~\ref{eq:sphereexp} becomes
\begin{align}
	P(\vec s) &= \frac{e^{-\textrm{tr}M/6}}{2^N}\left[ i_0\left( |\vec H| \right) - \frac{1}{2}\, i_2\left( |\vec H| \right) \hat H^{\mathsf T}\tilde M\hat H \right] + O\!\left( \tilde M^2 \right)
\end{align}
The two geometries are thus treated identically. Each is an expansion in the Bessel functions natural to its symmetry, with cylindrical harmonics and $I_n$ on the circle and spherical harmonics and $i_\ell$ on the sphere, and each collapses to its leading term when $M$ is isotropic.


In the isotropic case both distributions depend on $\vec s$ only through $|\vec H|$, and the interaction structure follows from expanding the logarithm, since
\begin{align}
	|\vec H|^2 &= \sum_{\rm i}|\vec h_{\rm i}|^2 + 2\sum_{\rm i<j}\left( \vec h_{\rm i}\cdot\vec h_{\rm j} \right)s_{\rm i}s_{\rm j}
\end{align}
holds exactly by $s_{\rm i}^2=1$. With $\log I_0(x) = x^2/4 - x^4/64 + \cdots$ the circle gives $J_{\rm ij} = \frac{1}{2}\vec h_{\rm i}\cdot\vec h_{\rm j}$ at leading order, and with $\log\left( \sinh x/x \right) = x^2/6 - x^4/180 + \cdots$ the sphere gives $J_{\rm ij} = \frac{1}{3}\vec h_{\rm i}\cdot\vec h_{\rm j}$. The quartic and higher terms supply the four-body and higher even-order interactions.

\begin{figure}
\centering
\includegraphics[width=\linewidth]{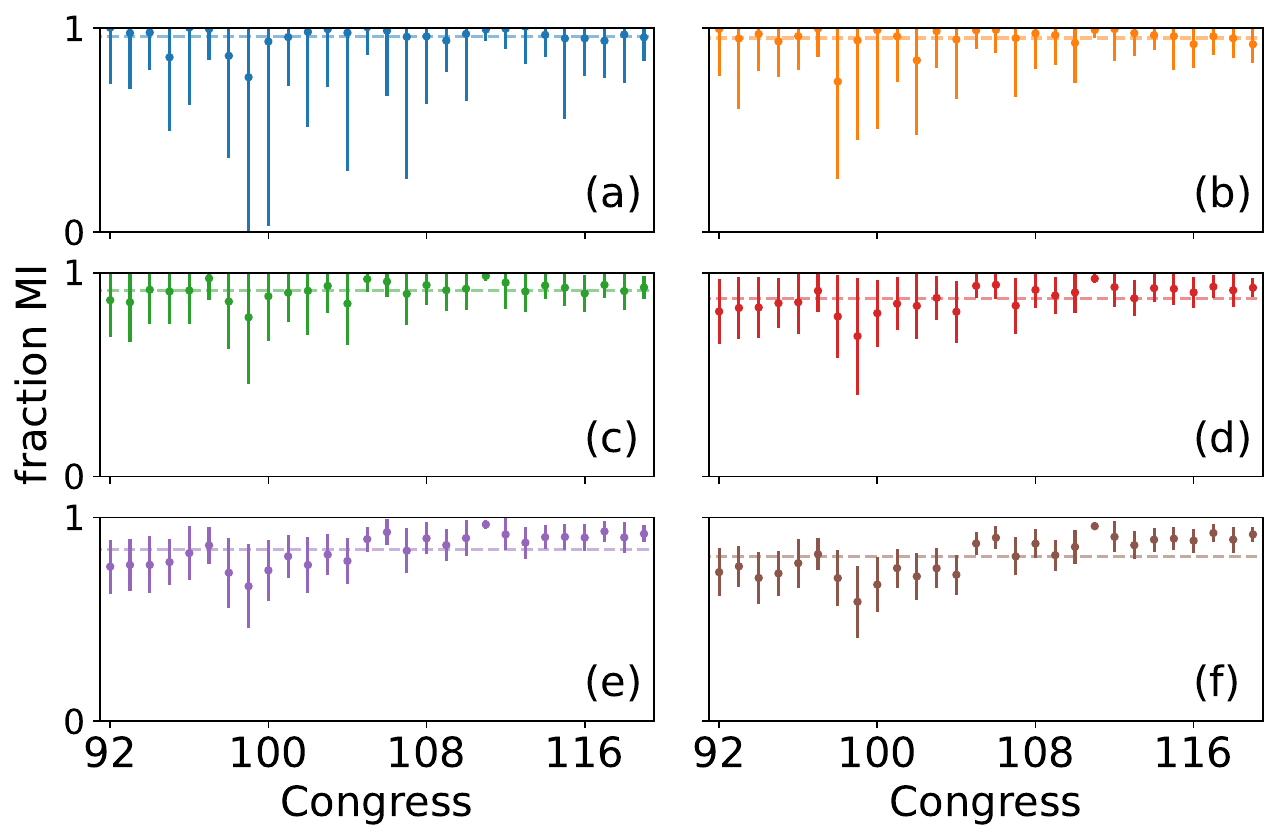}
\caption{Fraction of multi-information capture for odd-sized median groups $N=3$ through $N=15$ from left to right, top to bottom for $K=3$ model, except for $N=11$, which is shown in main text. Generally, the model does well and, importantly, it tends to do best when error bars are small and the state space is well-sampled.}
\end{figure}

\begin{figure}
\centering
\includegraphics[width=\linewidth]{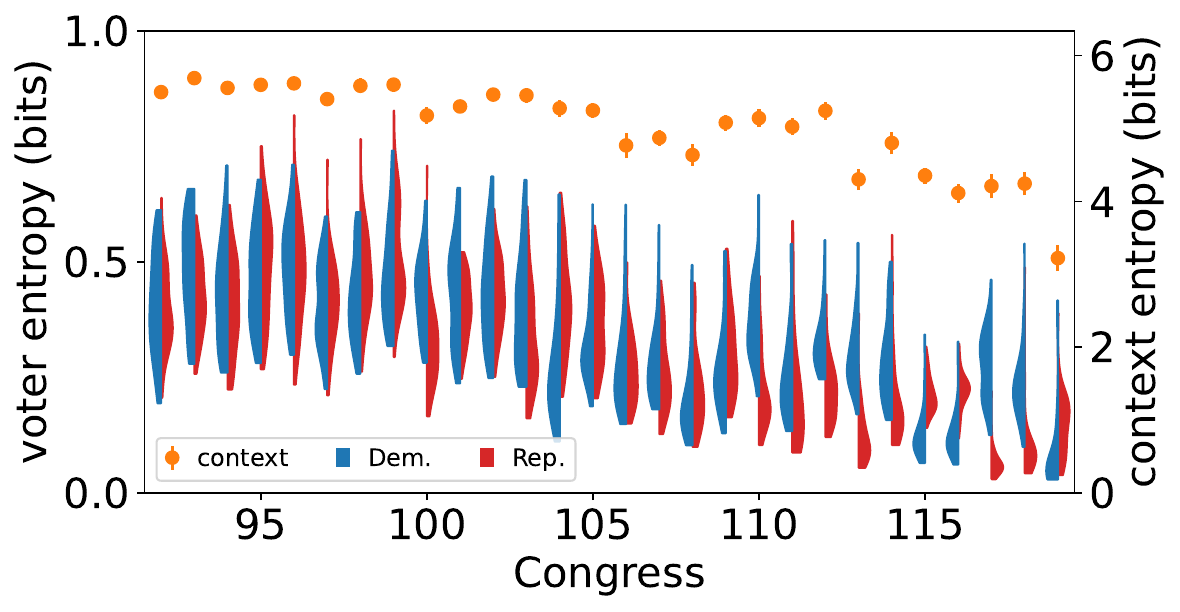}
\caption{Change in individual and collective entropies for the $K=4$ model as in Figure \ref{fig:complexity}.}
\end{figure}

\section{Voting preference maps}\label{si sec: maps}
We show each of the voting preference maps for all sessions from the \sessionn{92} through the \session{119} with the exception of the two that are discussed in the main text in Figure~\ref{fig:maps}. We highlight some of the voters that we discuss specifically in the main text.

\begin{figure*}[p!]
\centering
	\includegraphics[width=.49\linewidth]{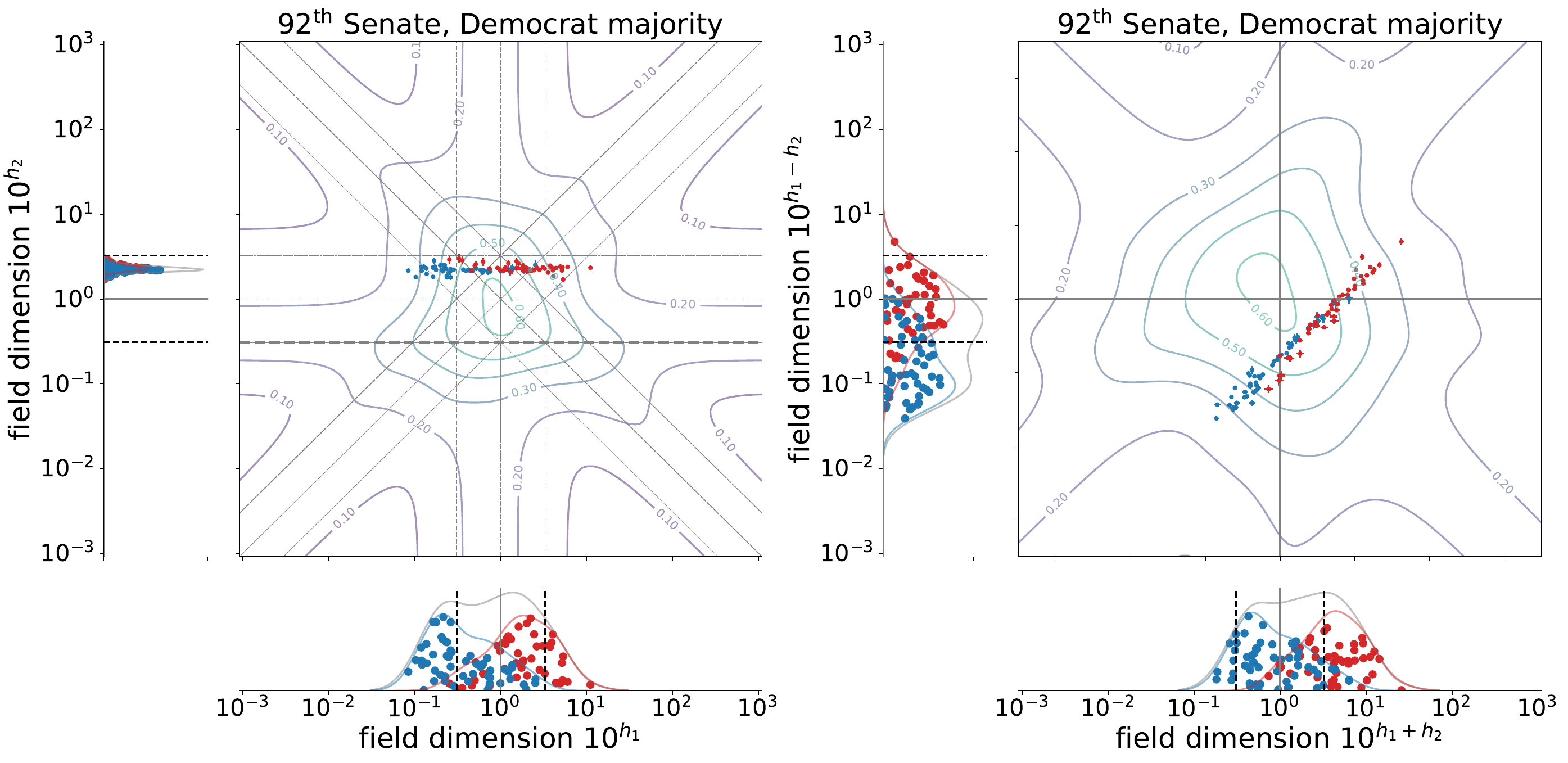}\includegraphics[width=.49\linewidth]{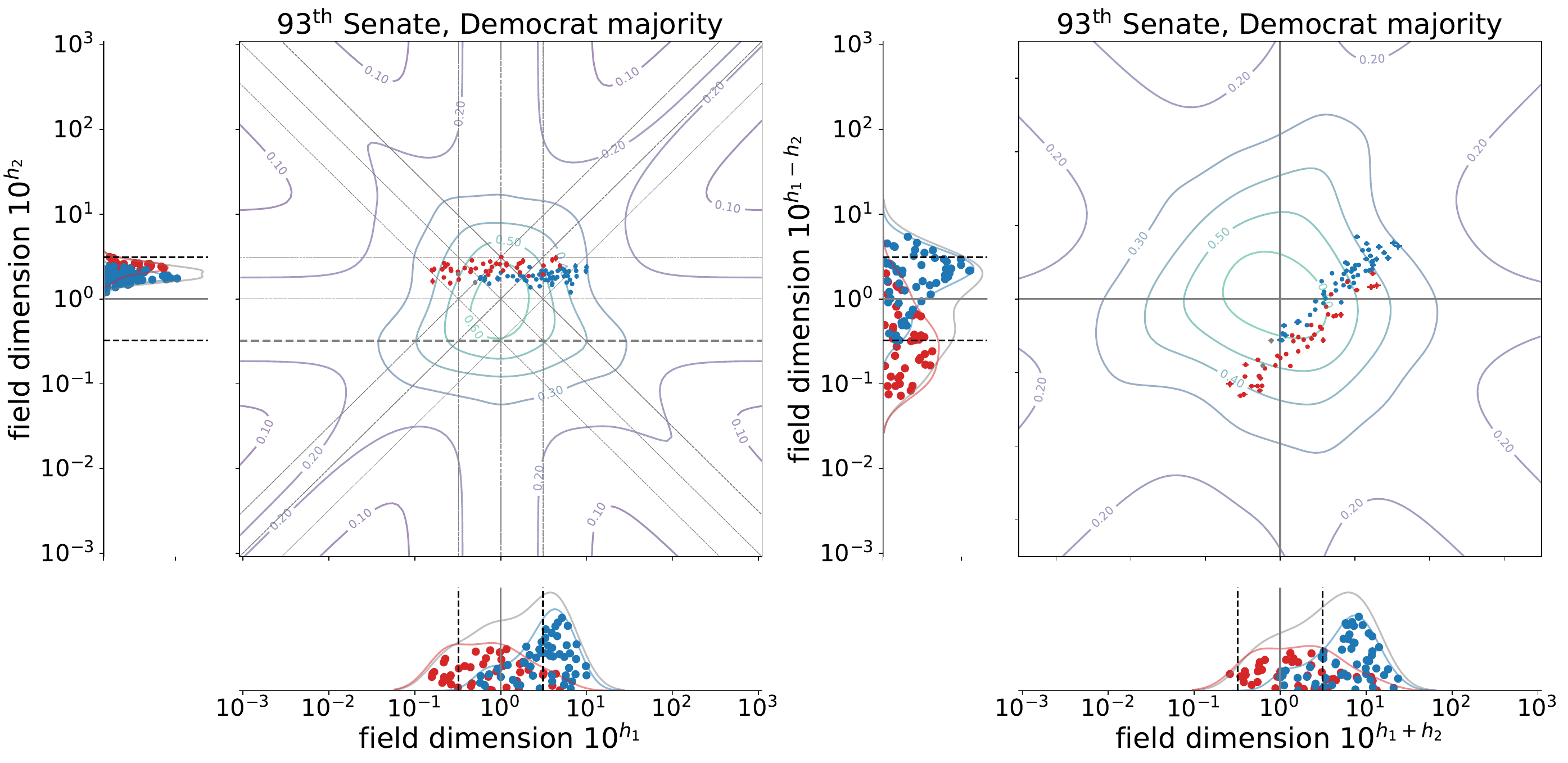}\\
	\includegraphics[width=.49\linewidth]{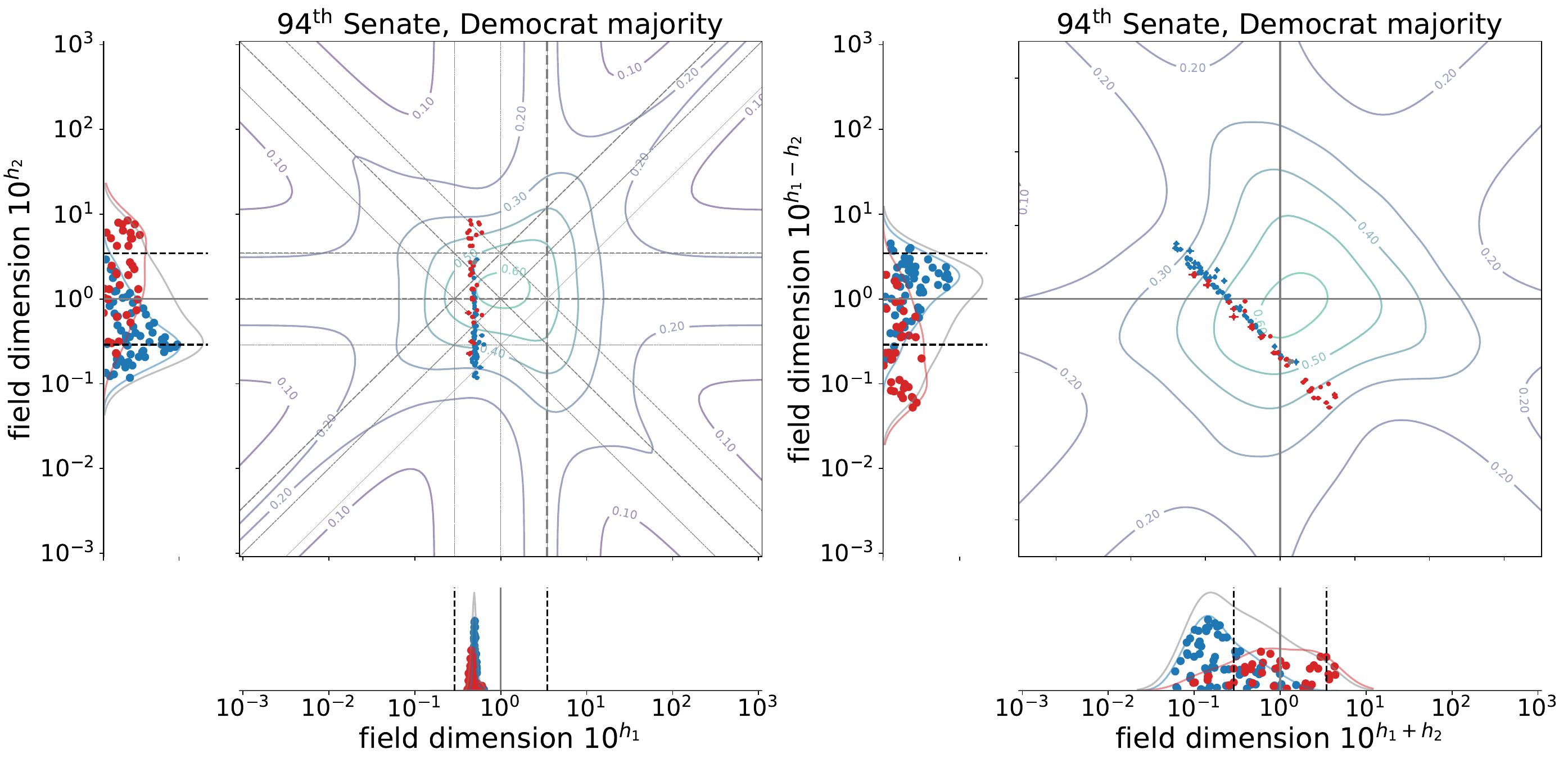}\includegraphics[width=.49\linewidth]{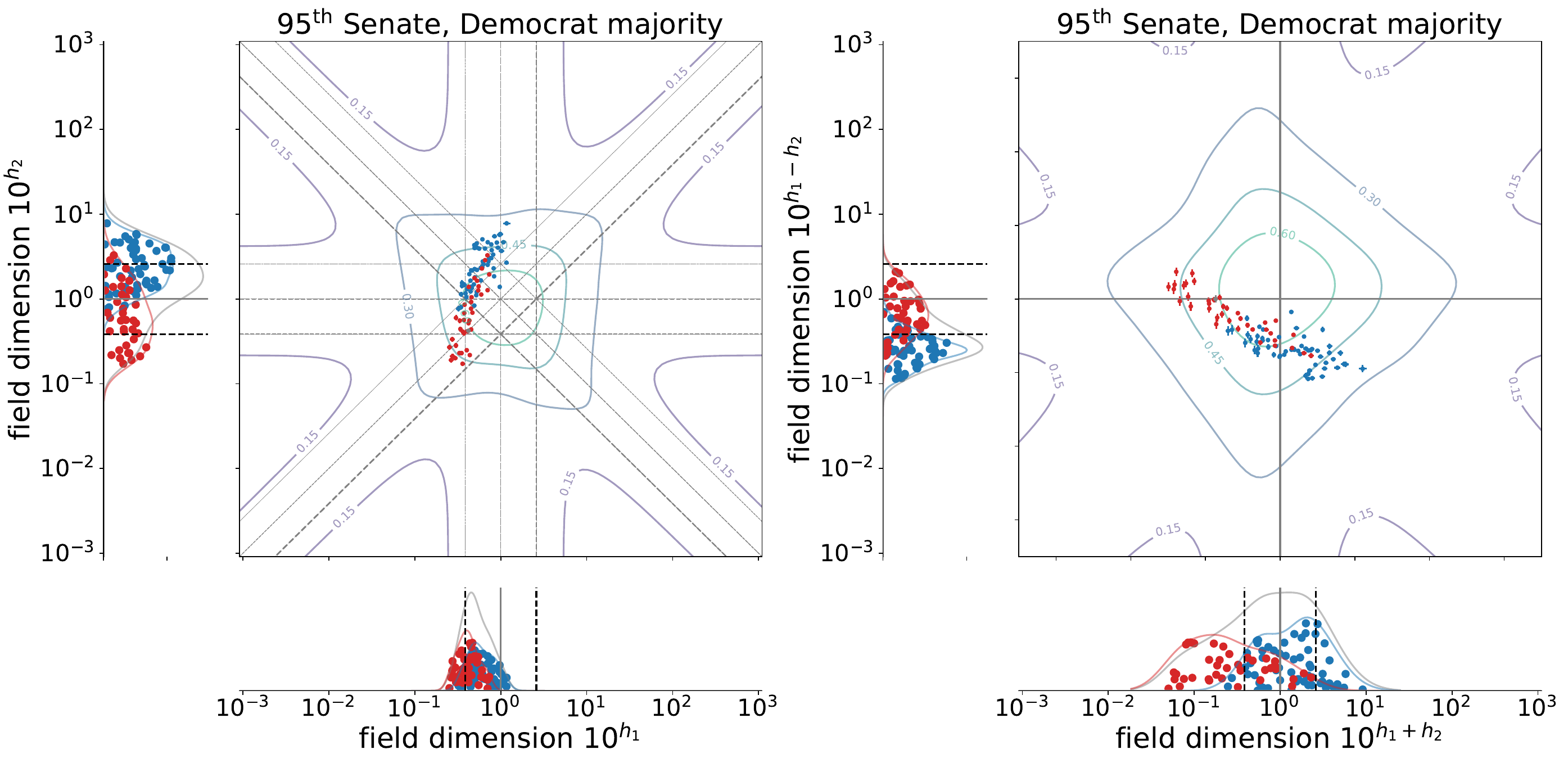}\\
	\includegraphics[width=.49\linewidth]{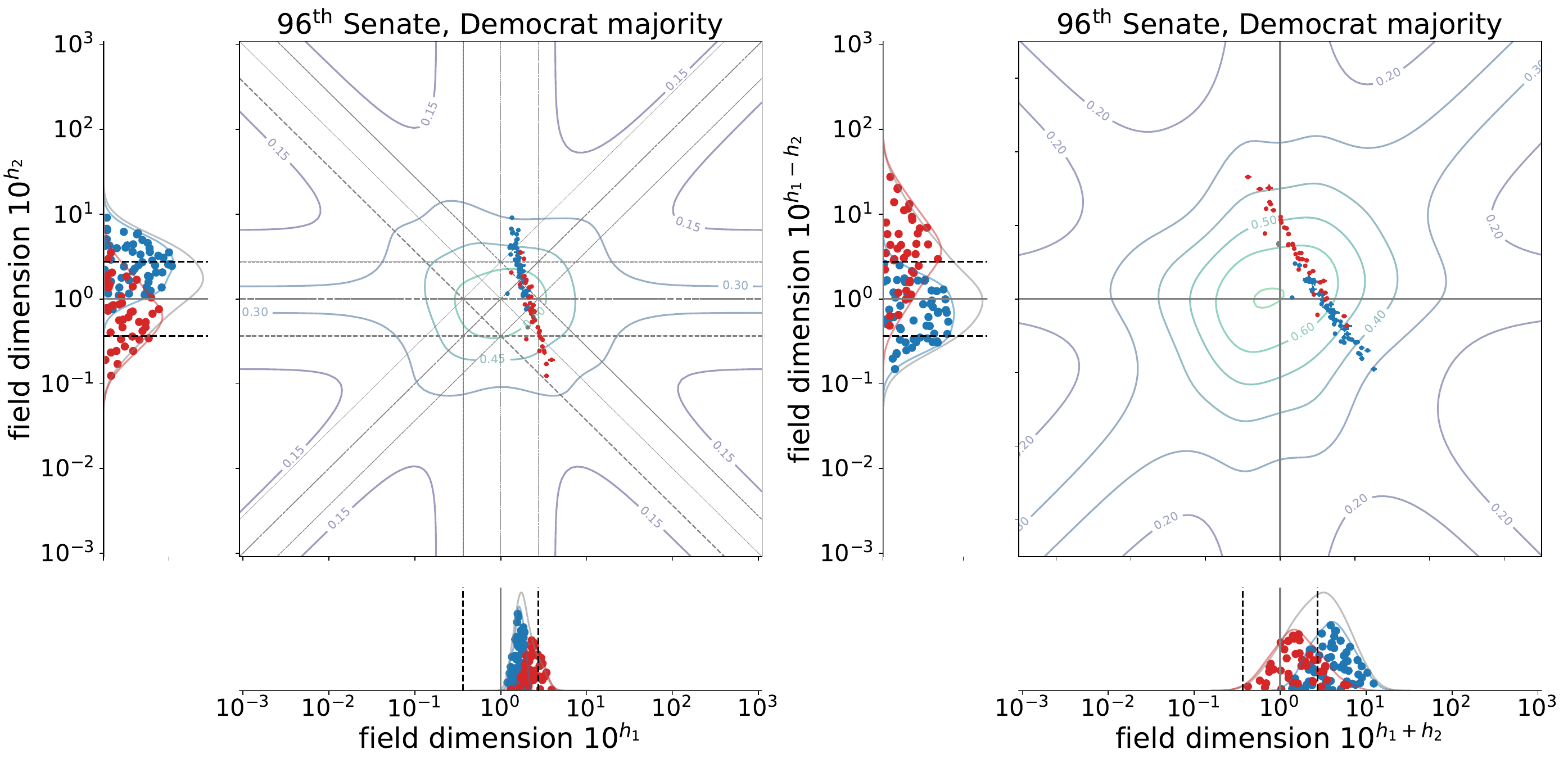}\includegraphics[width=.49\linewidth]{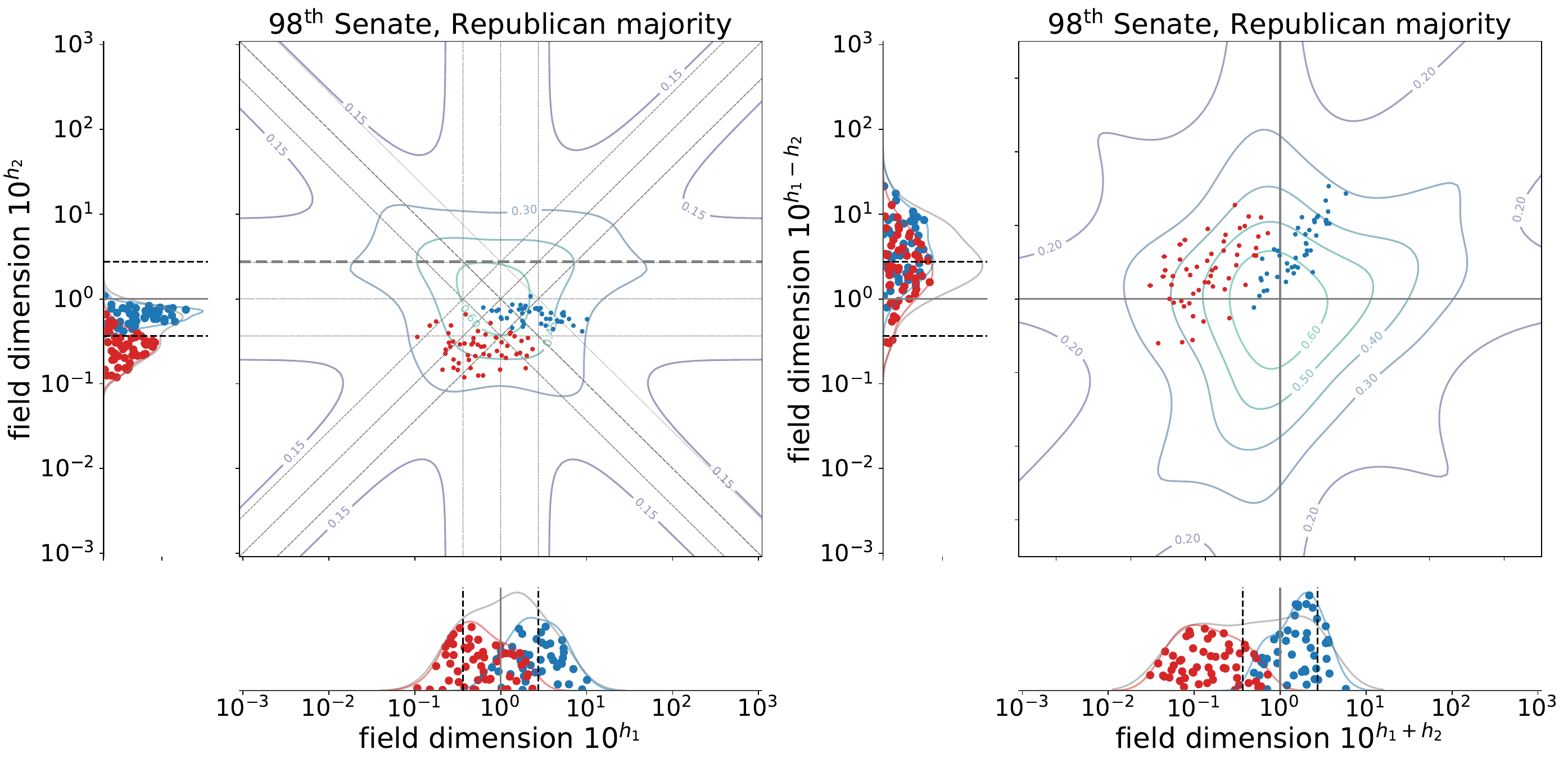}\\
	\includegraphics[width=.49\linewidth]{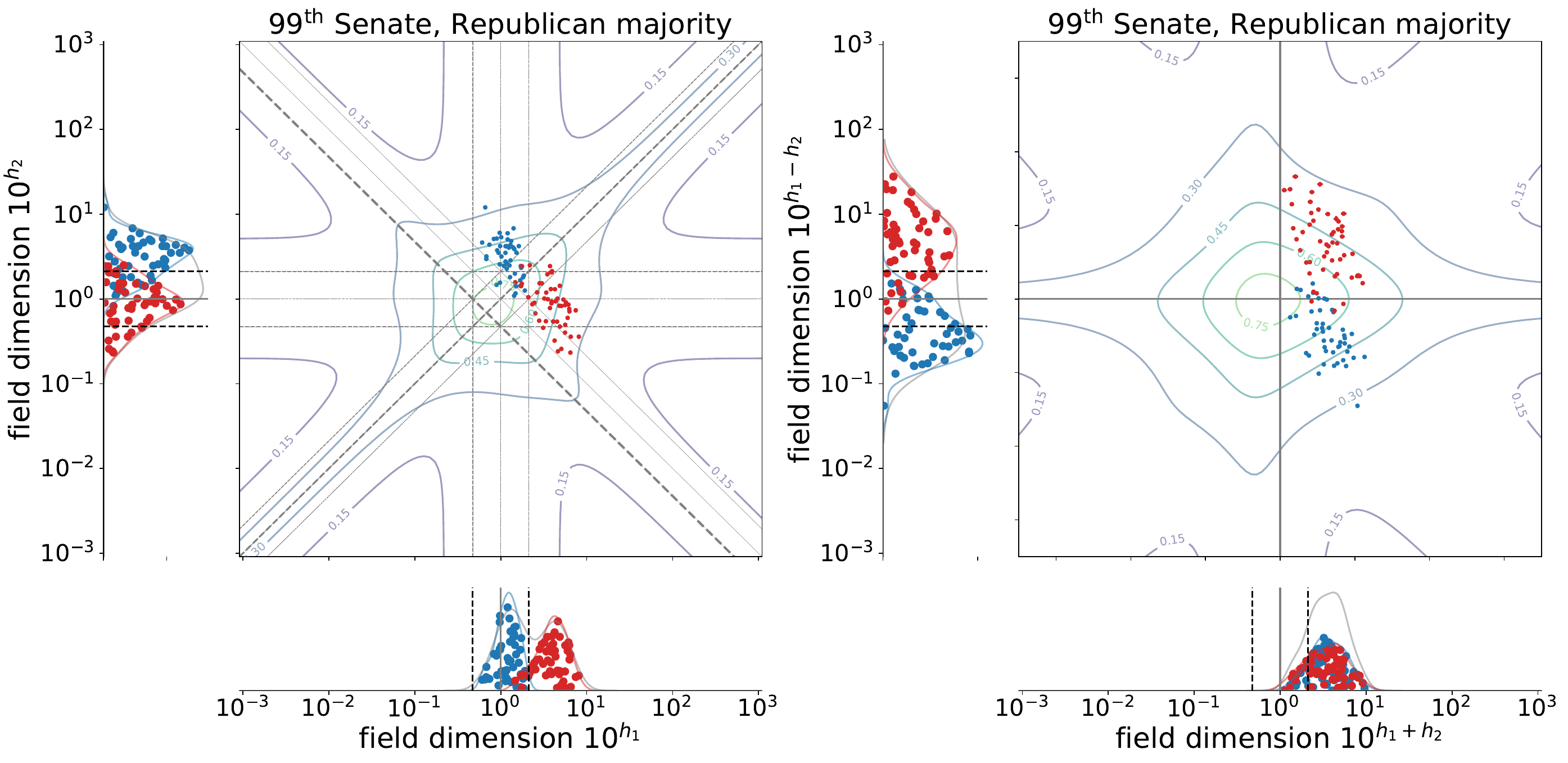}\includegraphics[width=.49\linewidth]{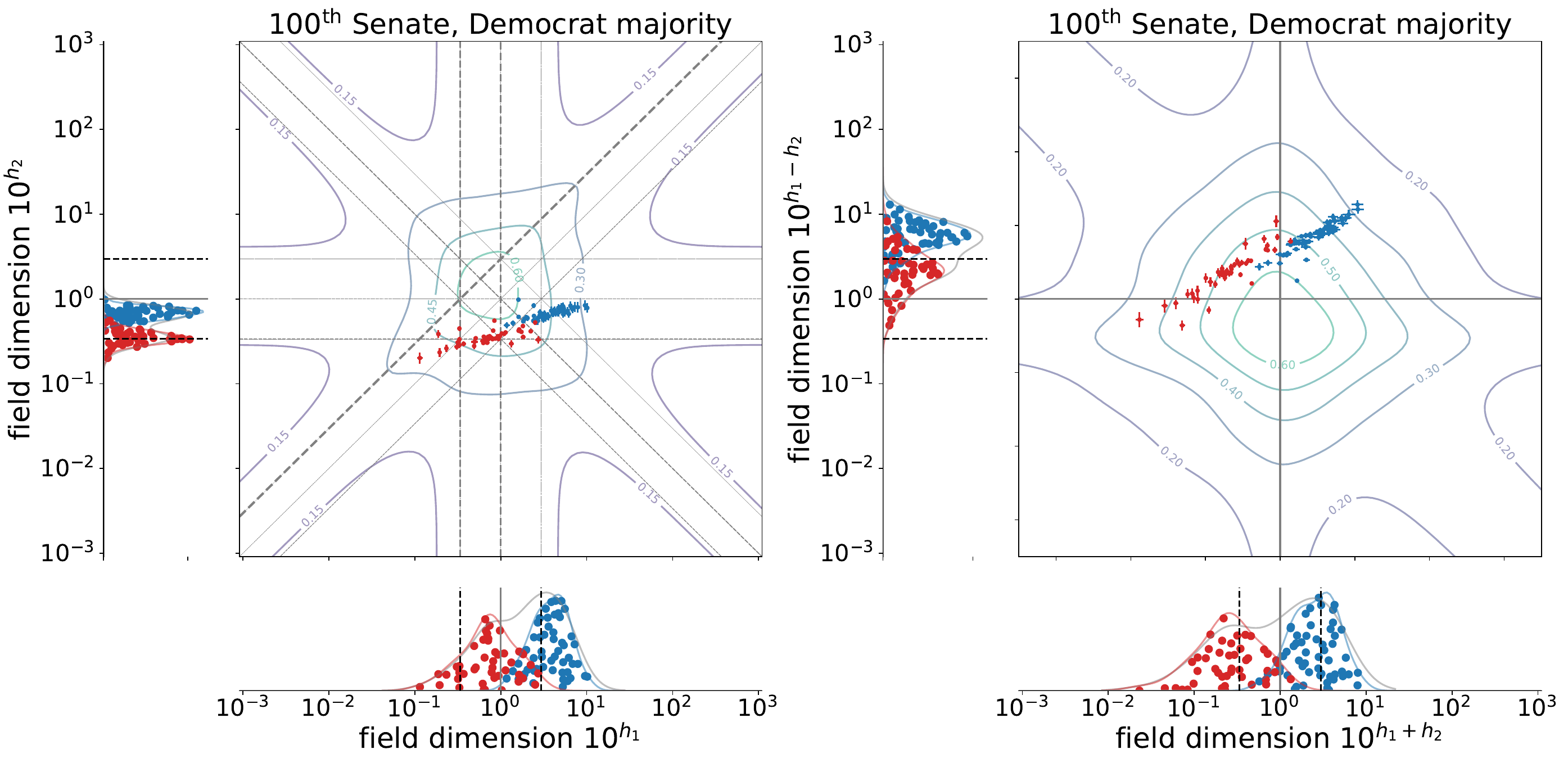}\\
	\includegraphics[width=.49\linewidth]{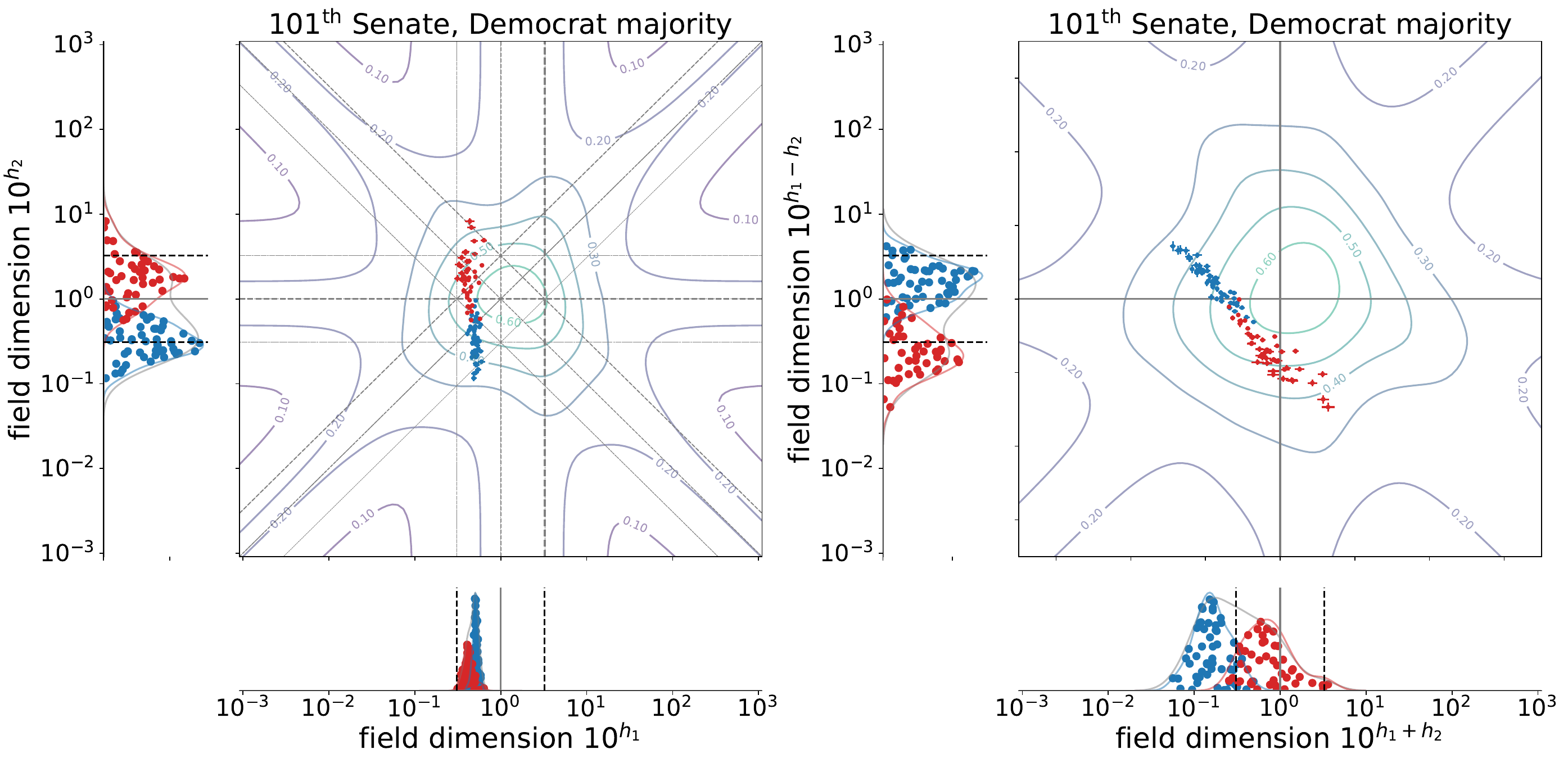}\includegraphics[width=.49\linewidth]{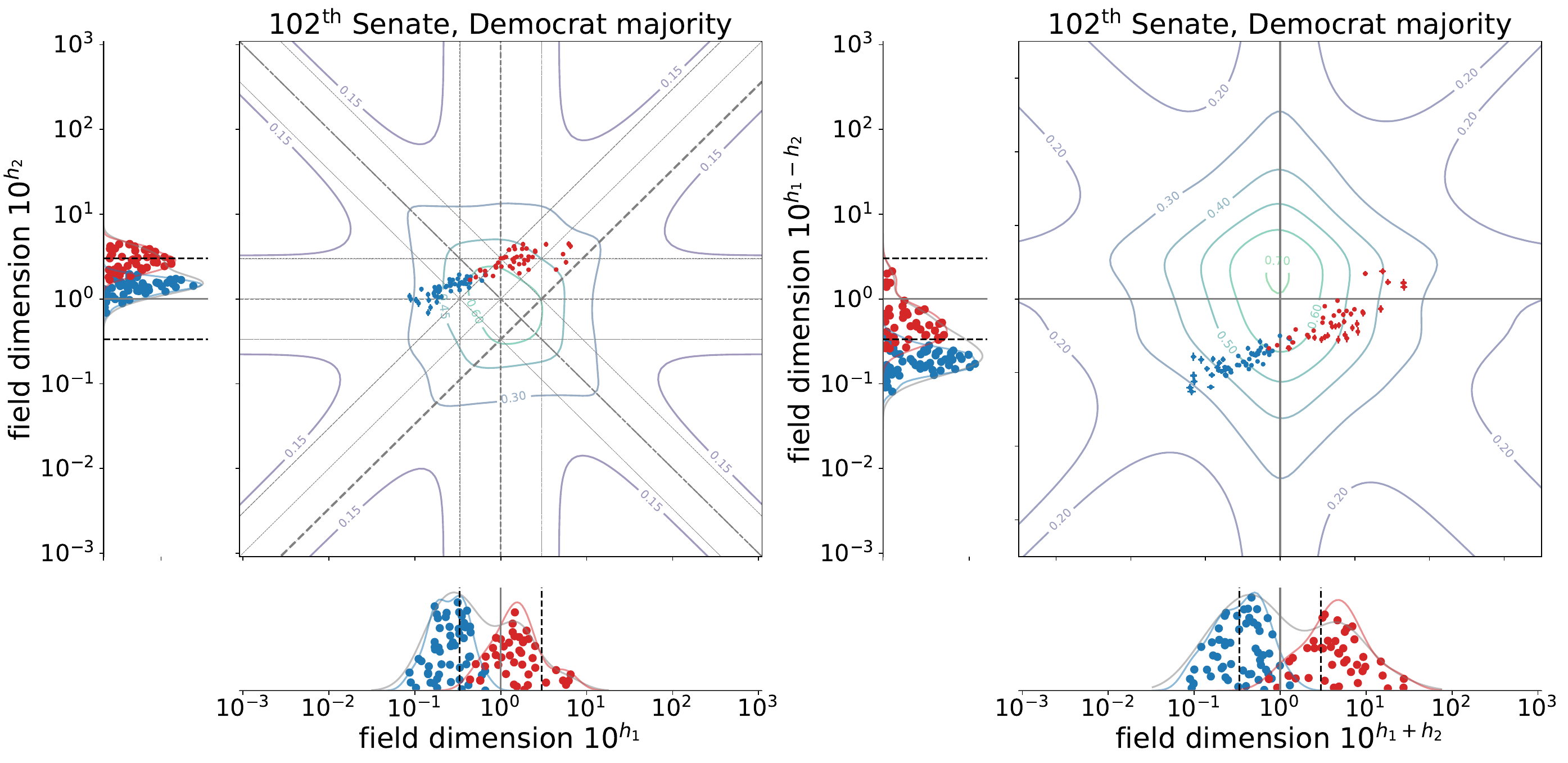}
	\caption{Parameter projection for $K=3$. \sessionn{92} through \session{102} sessions, excluding the \session{97}. Each pair of panels shows the original projection and then the rotated projection.}\label{fig:maps2}
\end{figure*}

\begin{figure*}[p!]
\centering
	\includegraphics[width=.49\linewidth]{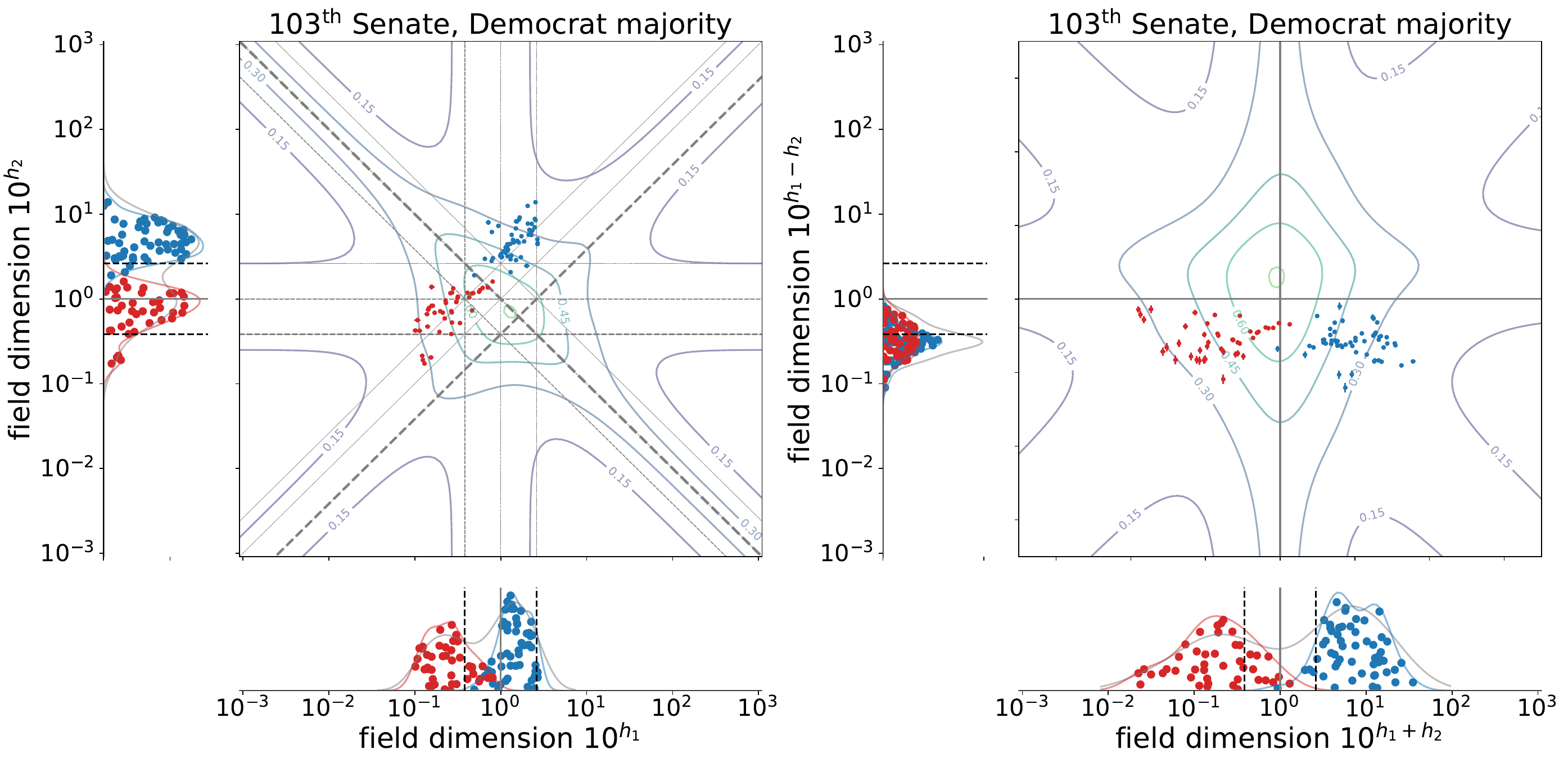}\includegraphics[width=.49\linewidth]{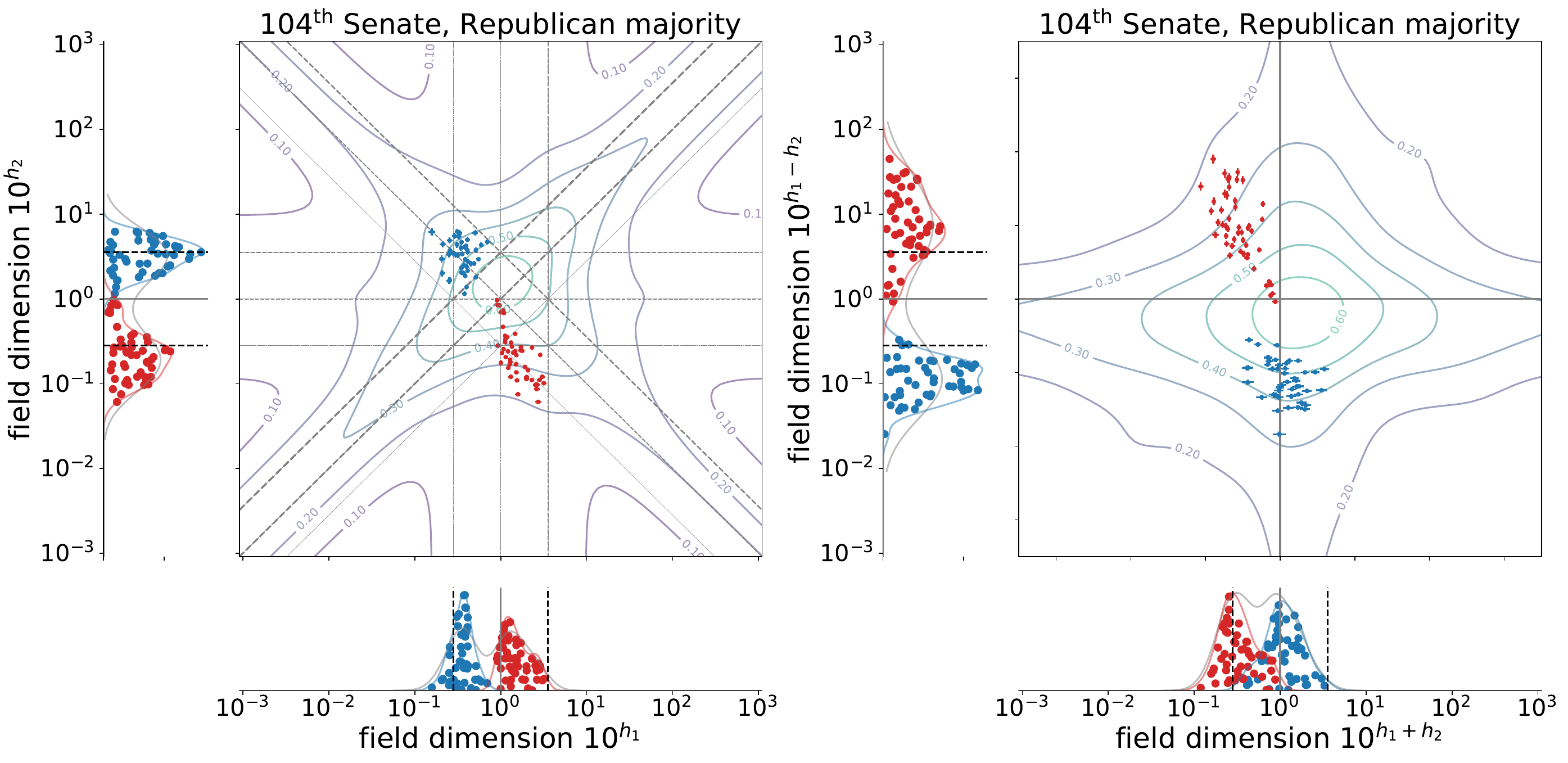}\\
	\includegraphics[width=.49\linewidth]{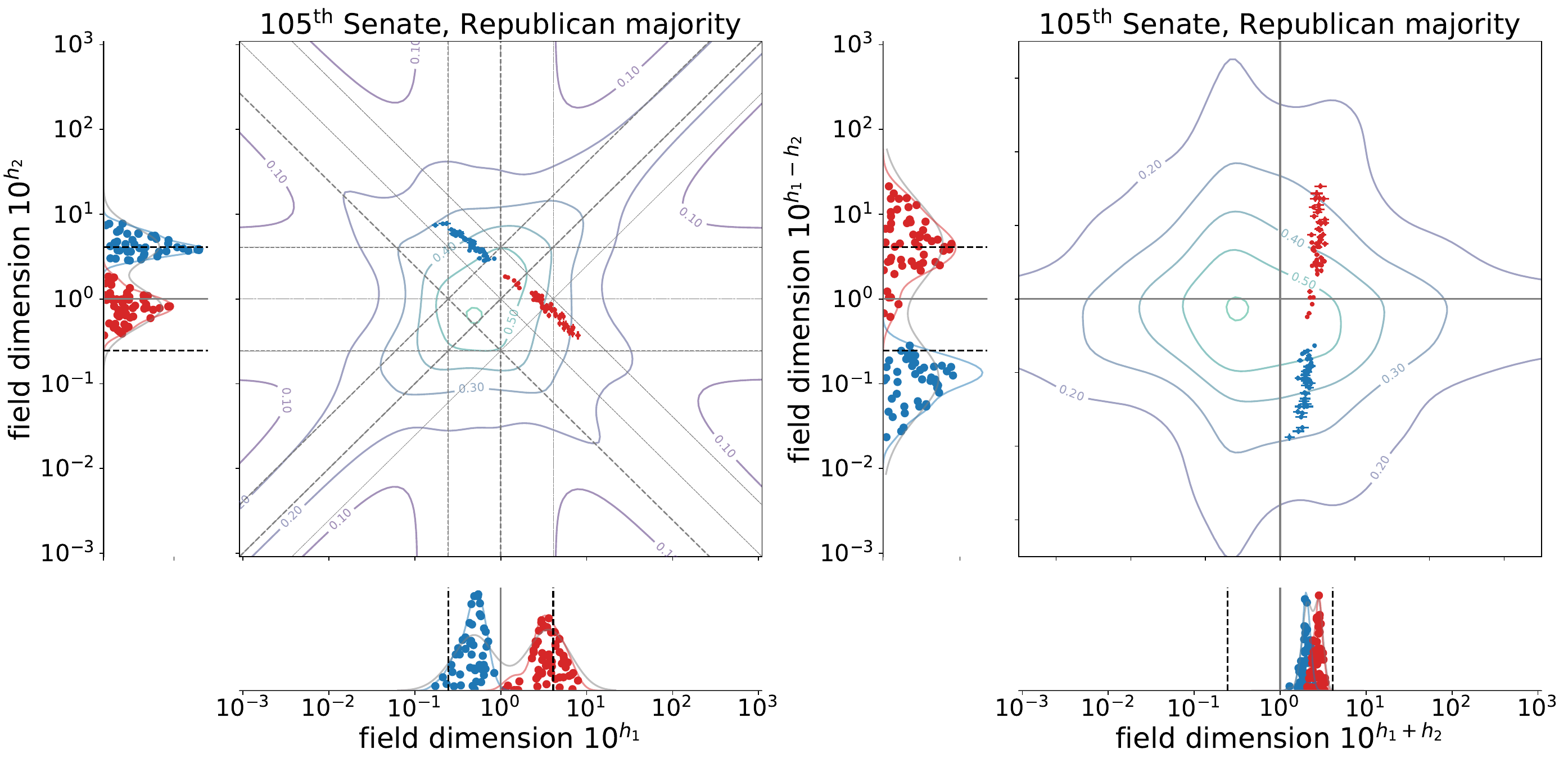}\includegraphics[width=.49\linewidth]{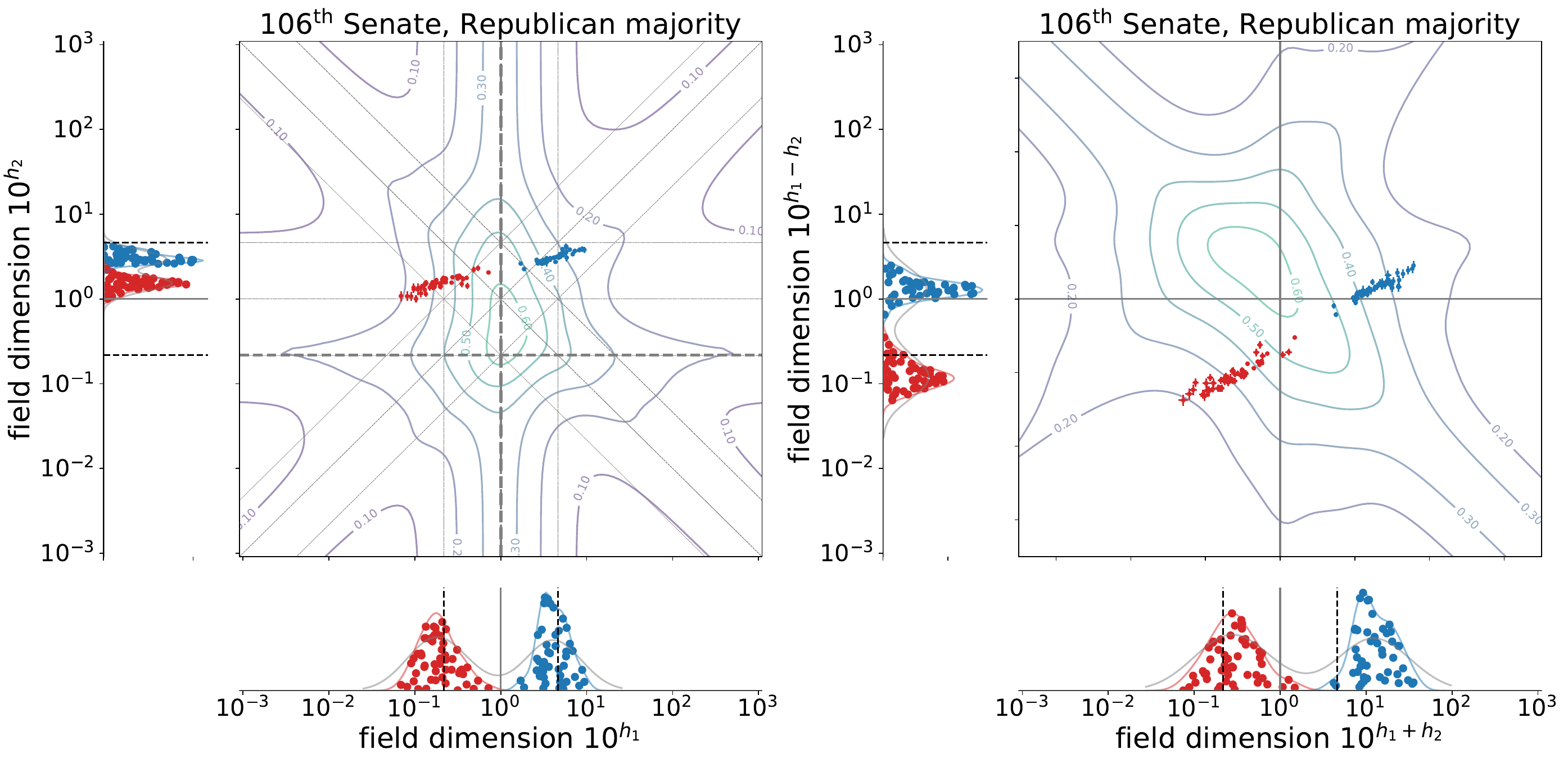}\\
	\includegraphics[width=.49\linewidth]{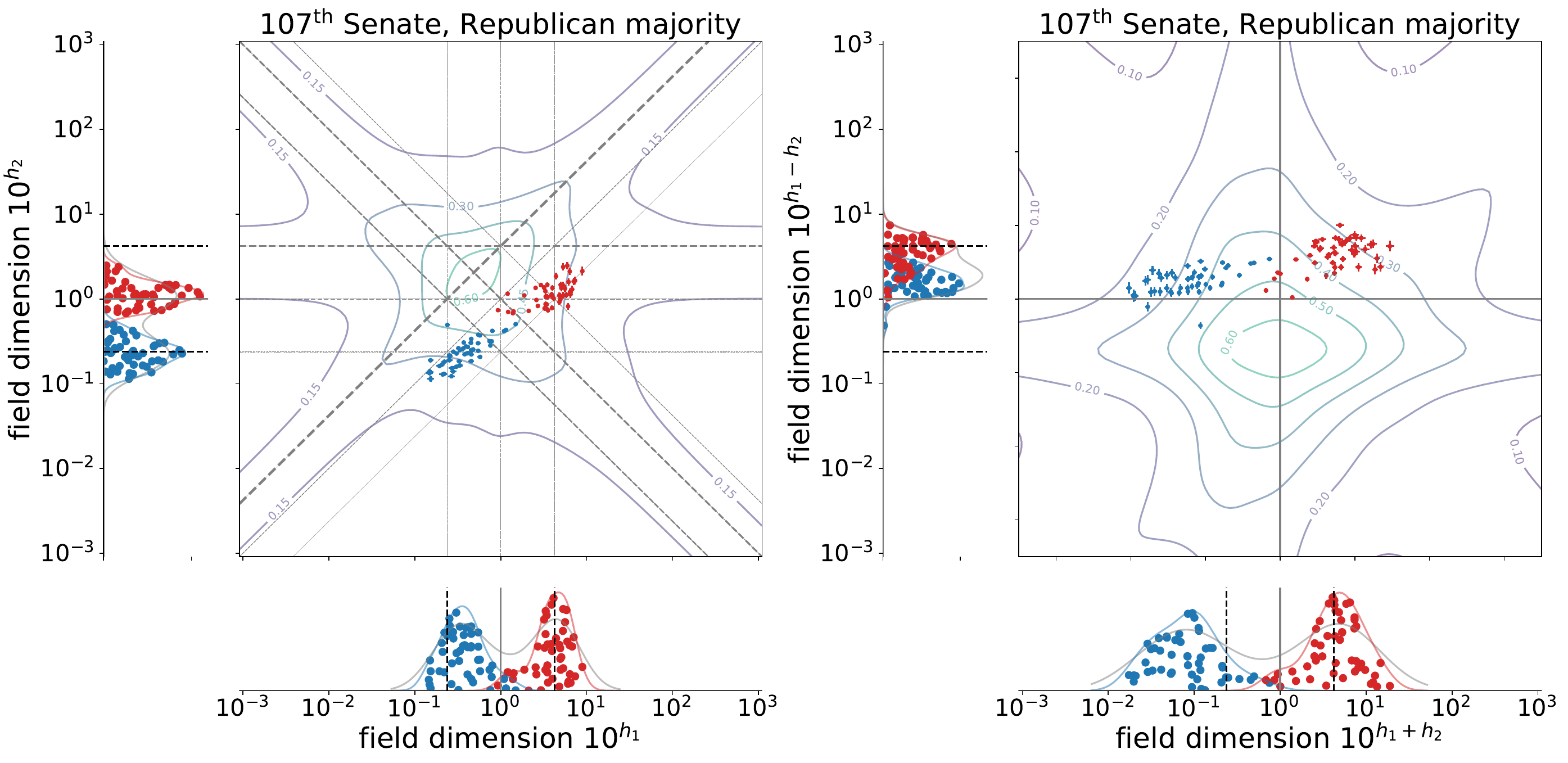}\includegraphics[width=.49\linewidth]{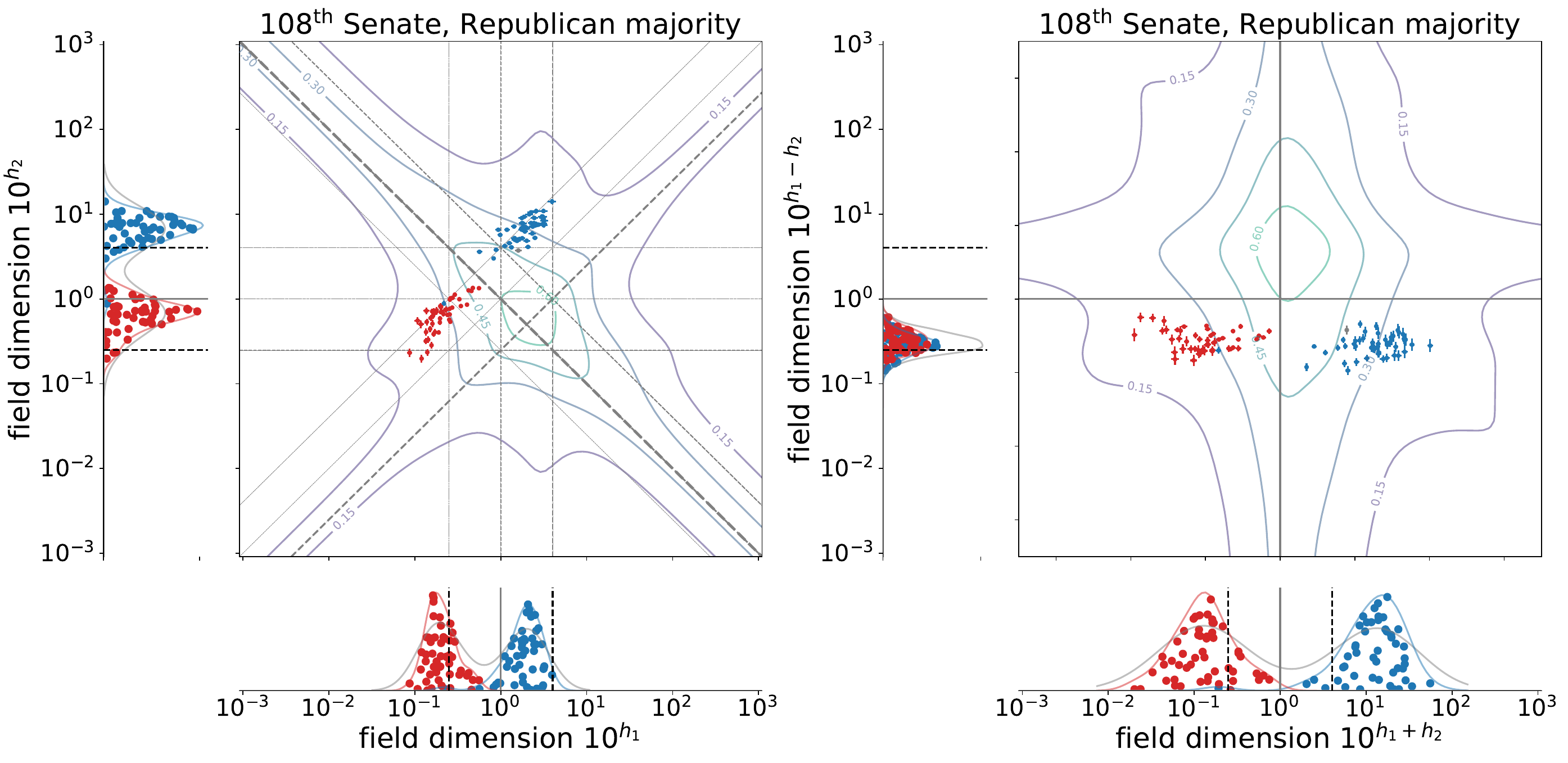}\\
	\includegraphics[width=.49\linewidth]{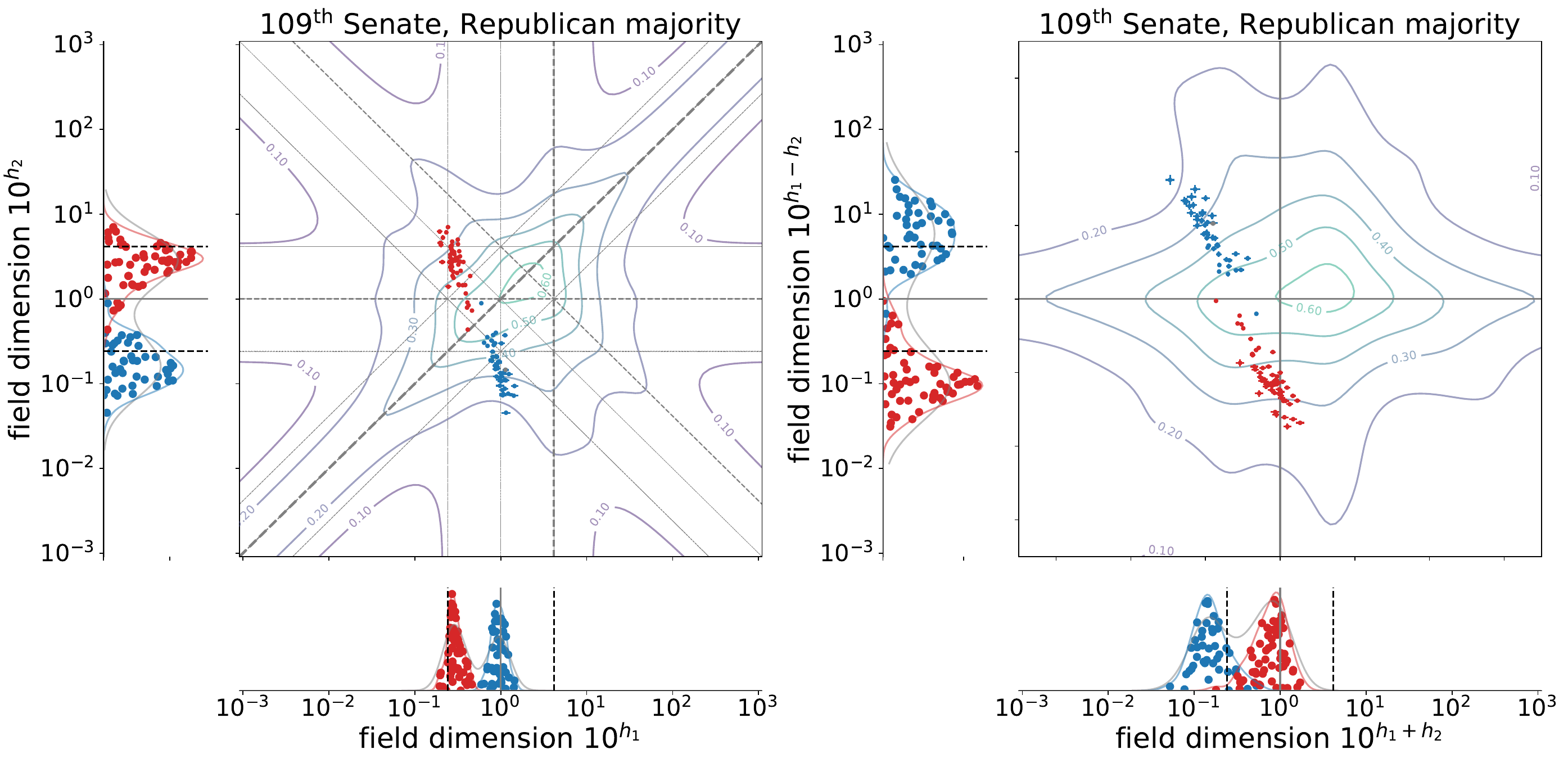}\includegraphics[width=.49\linewidth]{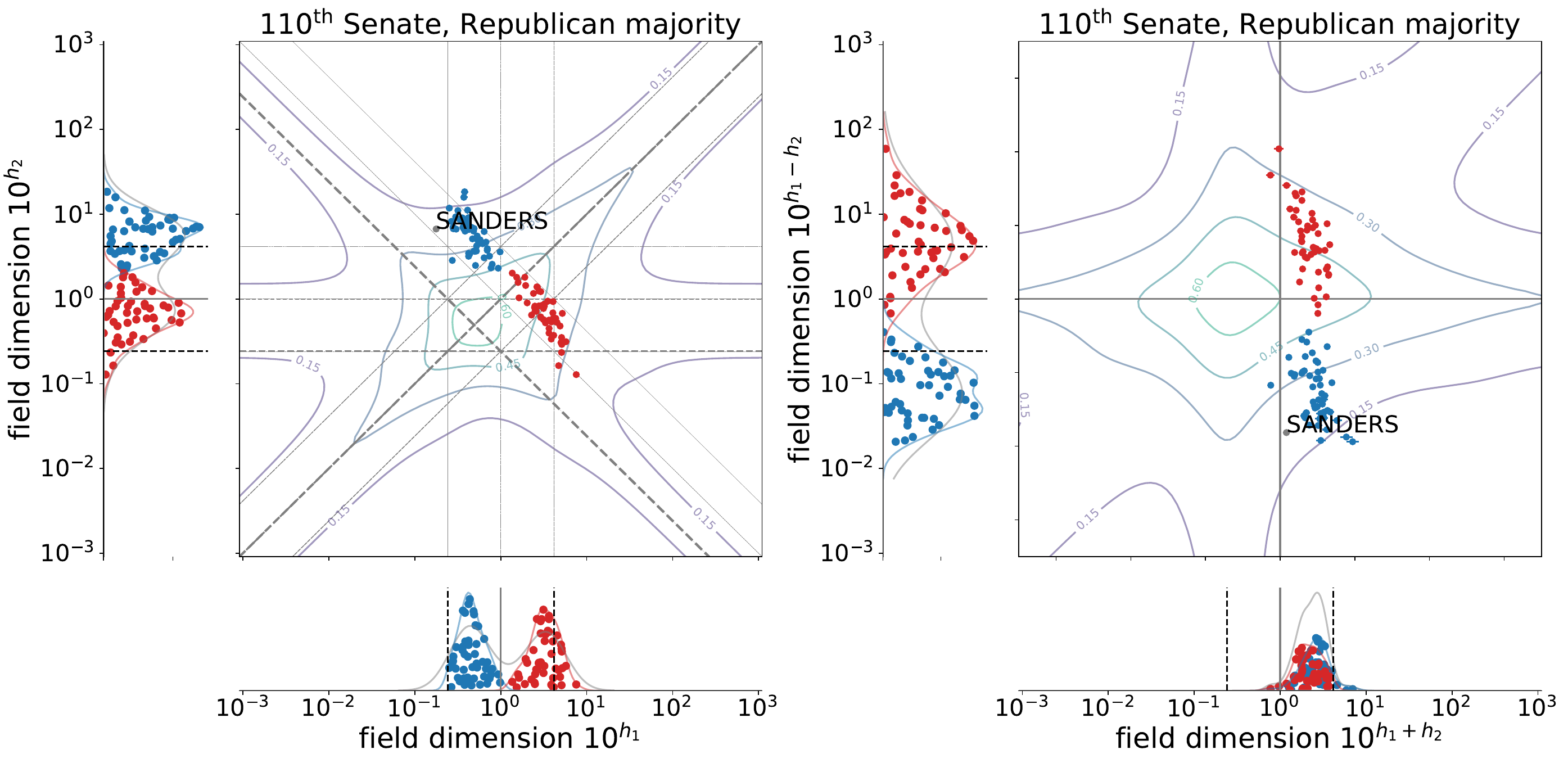}\\
	\includegraphics[width=.49\linewidth]{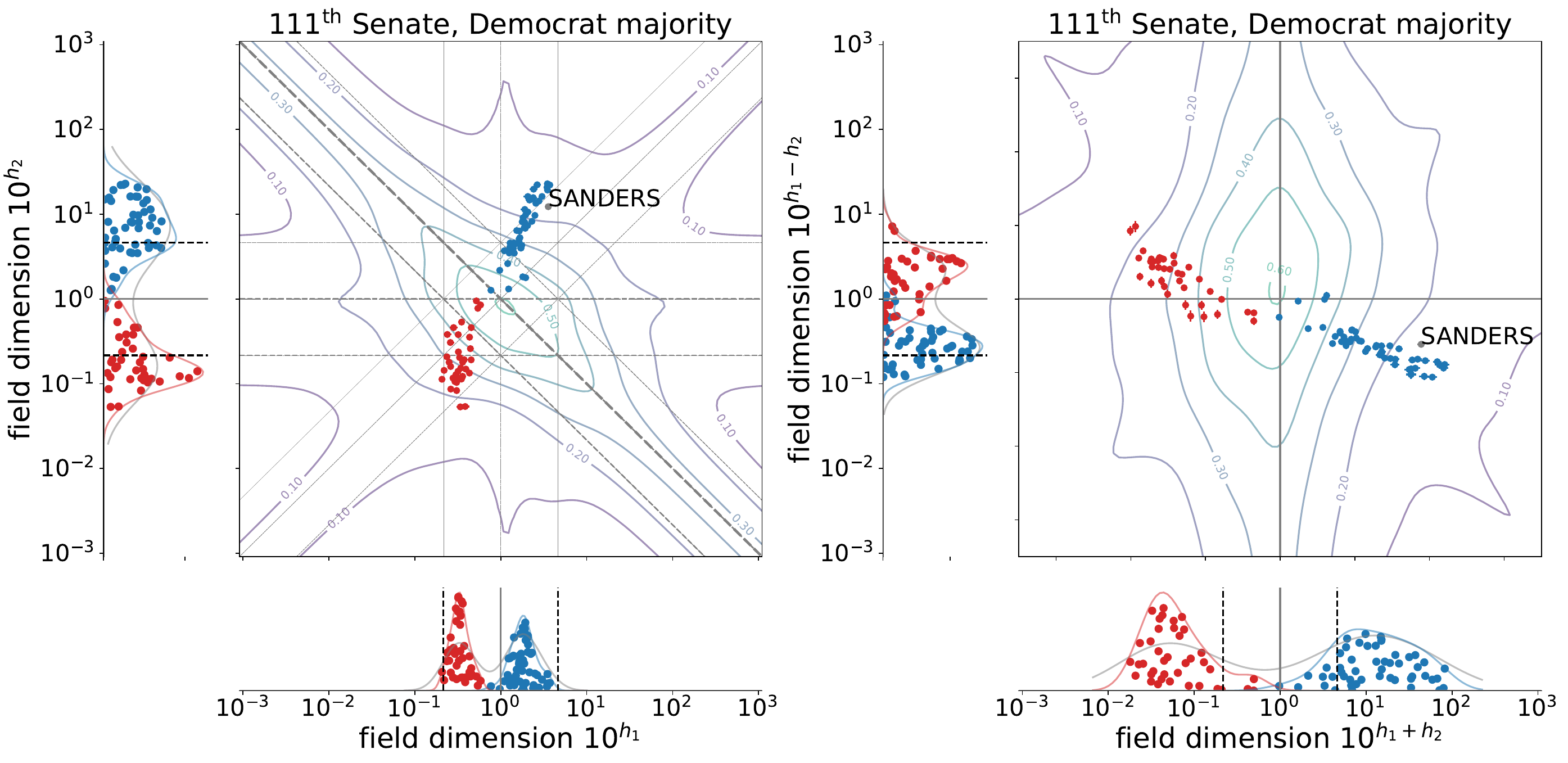}\includegraphics[width=.49\linewidth]{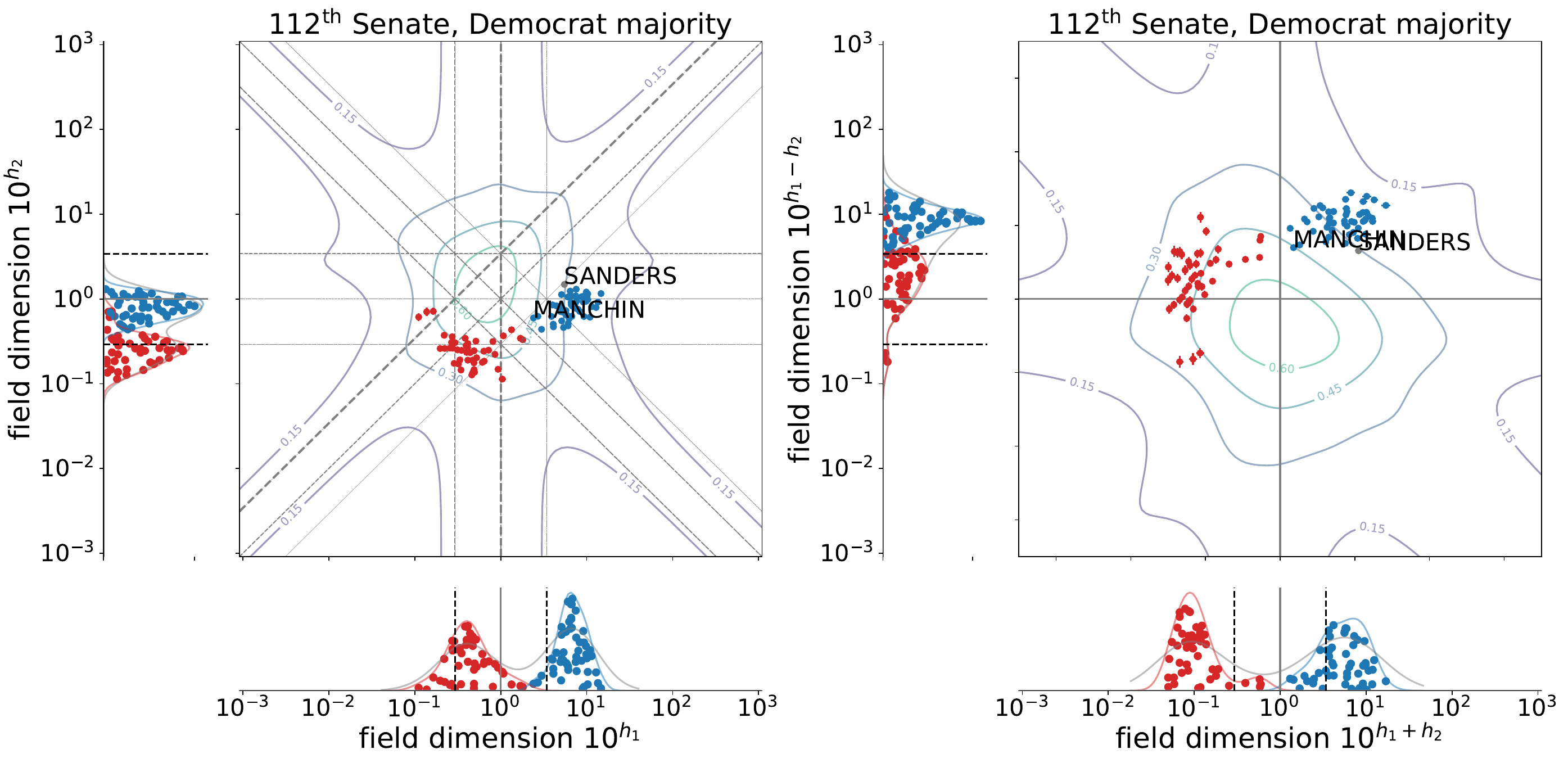}
	\caption{Parameter projection for $K=3$. \session{103} through \session{112} sessions.}\label{fig:maps3}
\end{figure*}

\begin{figure*}[p!]
\centering
	\includegraphics[width=.49\linewidth]{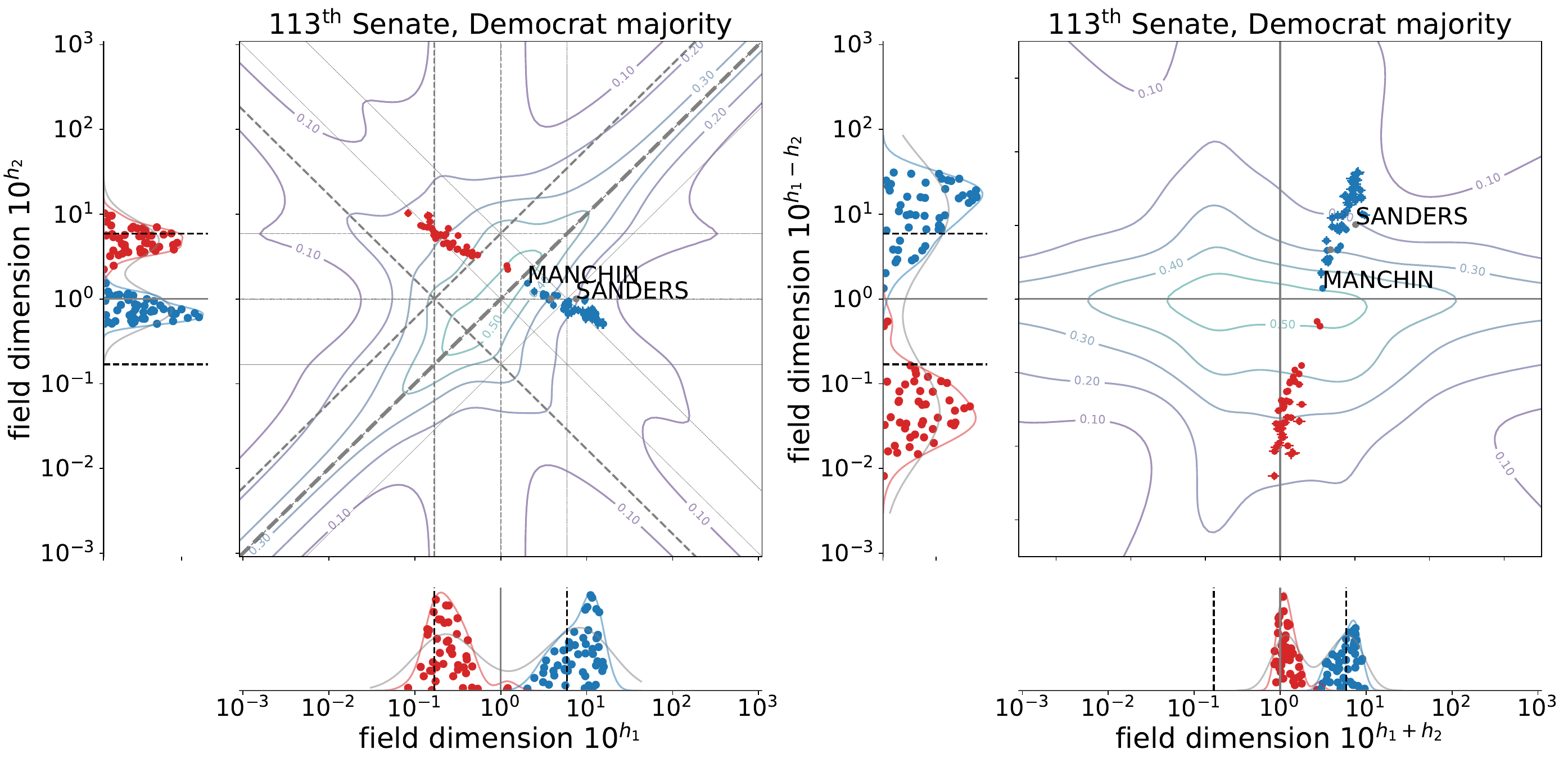}\includegraphics[width=.49\linewidth]{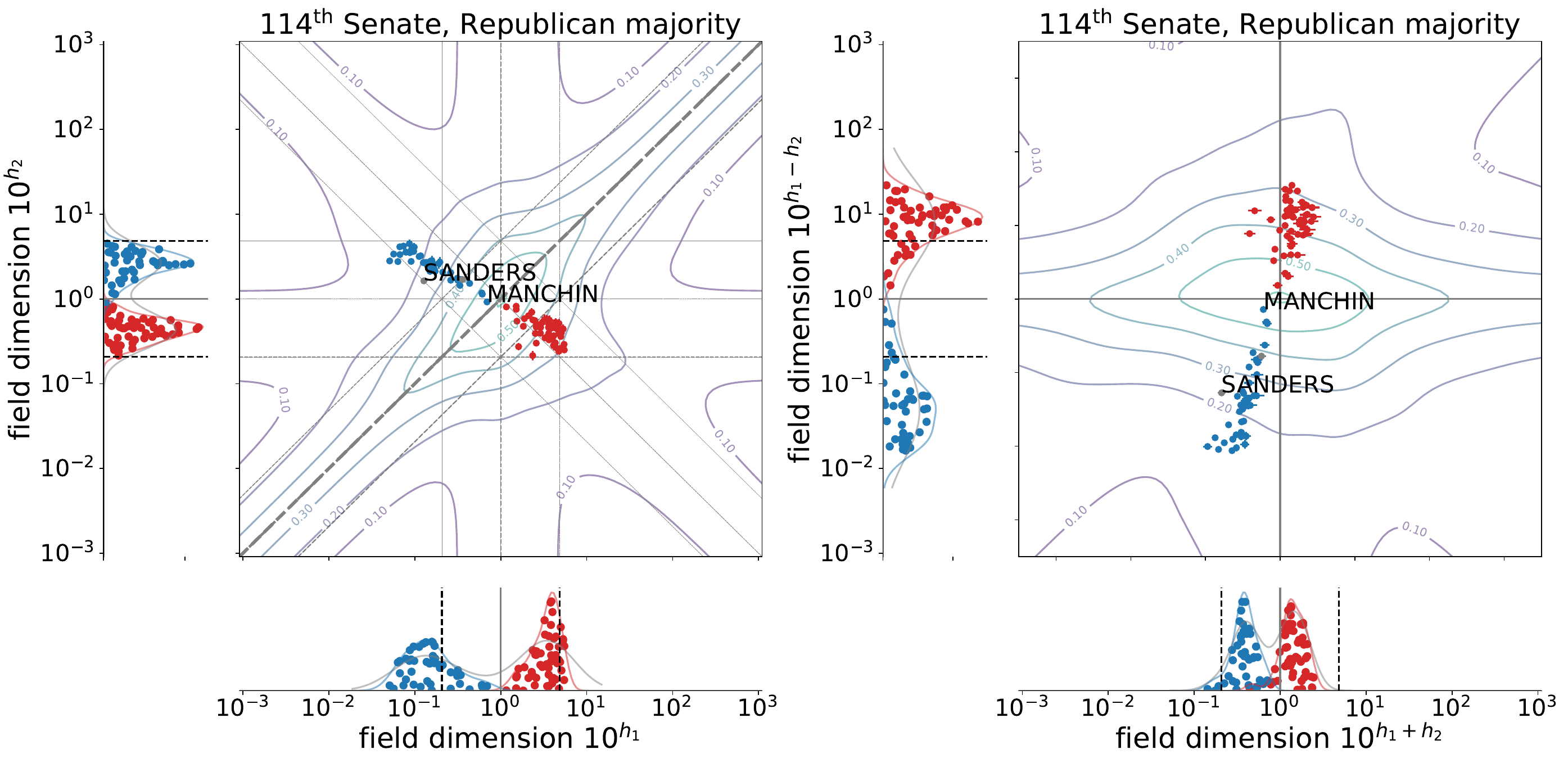}\\
	\includegraphics[width=.49\linewidth]{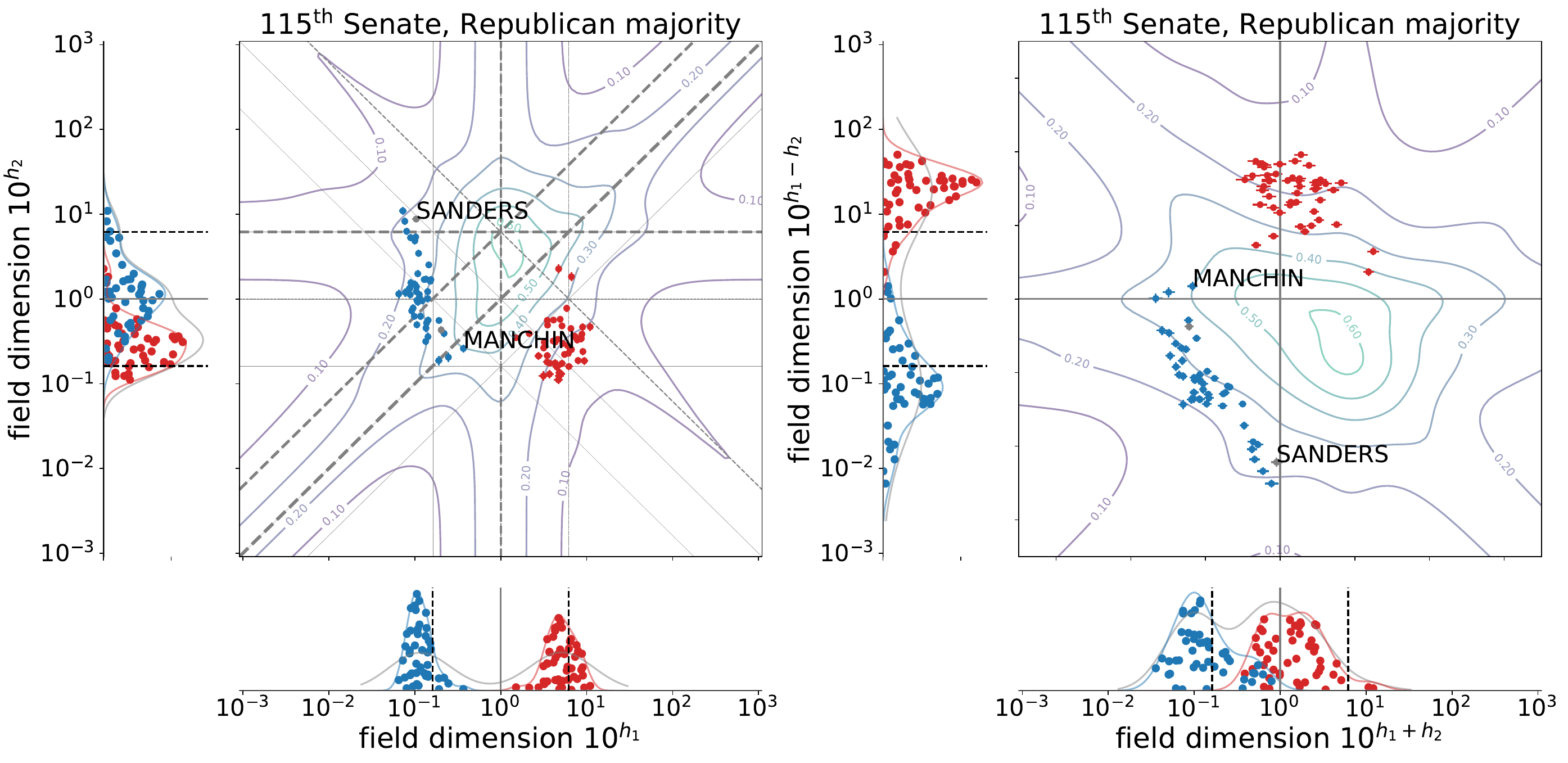}\includegraphics[width=.49\linewidth]{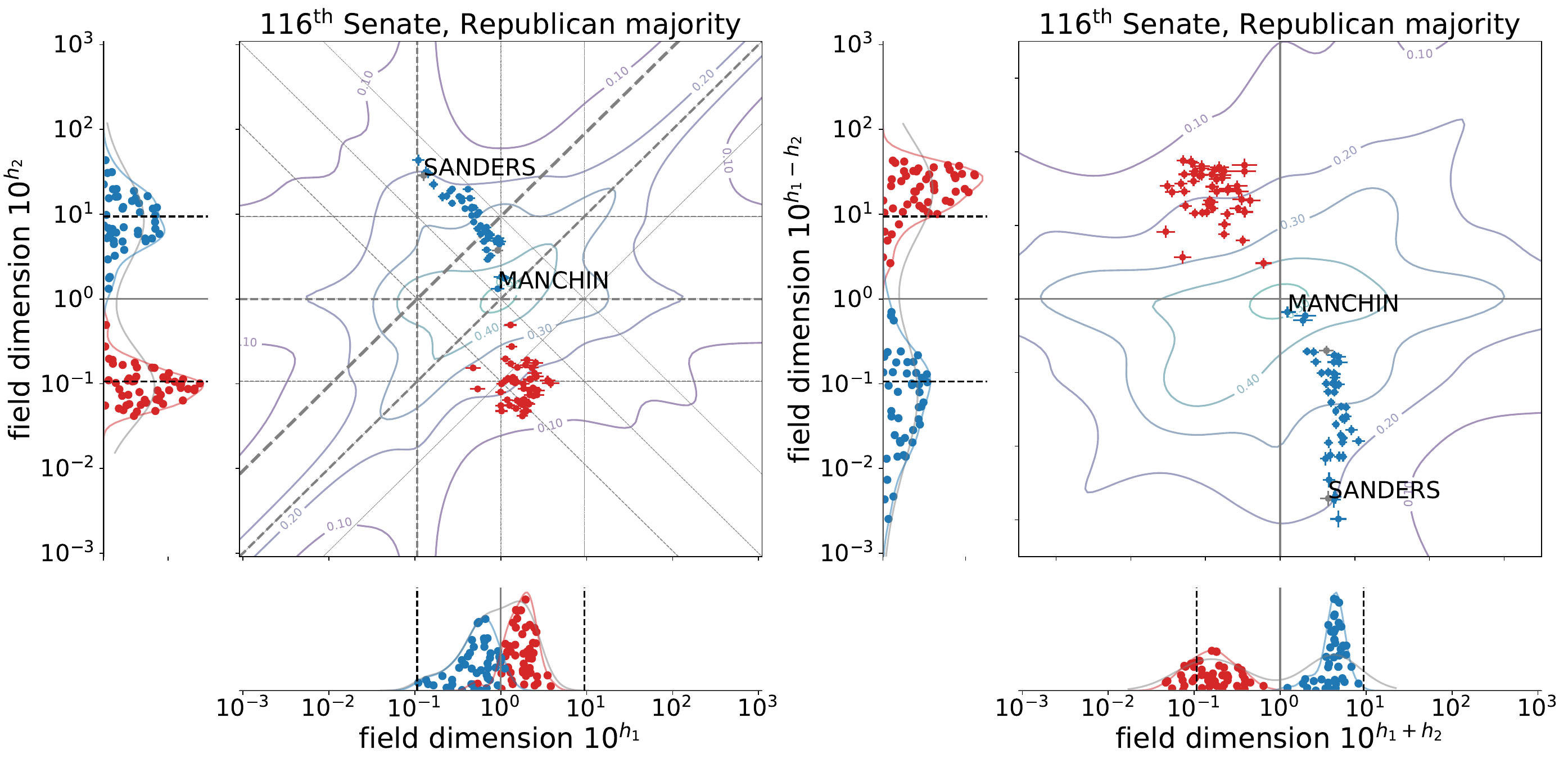}\\
	\includegraphics[width=.49\linewidth]{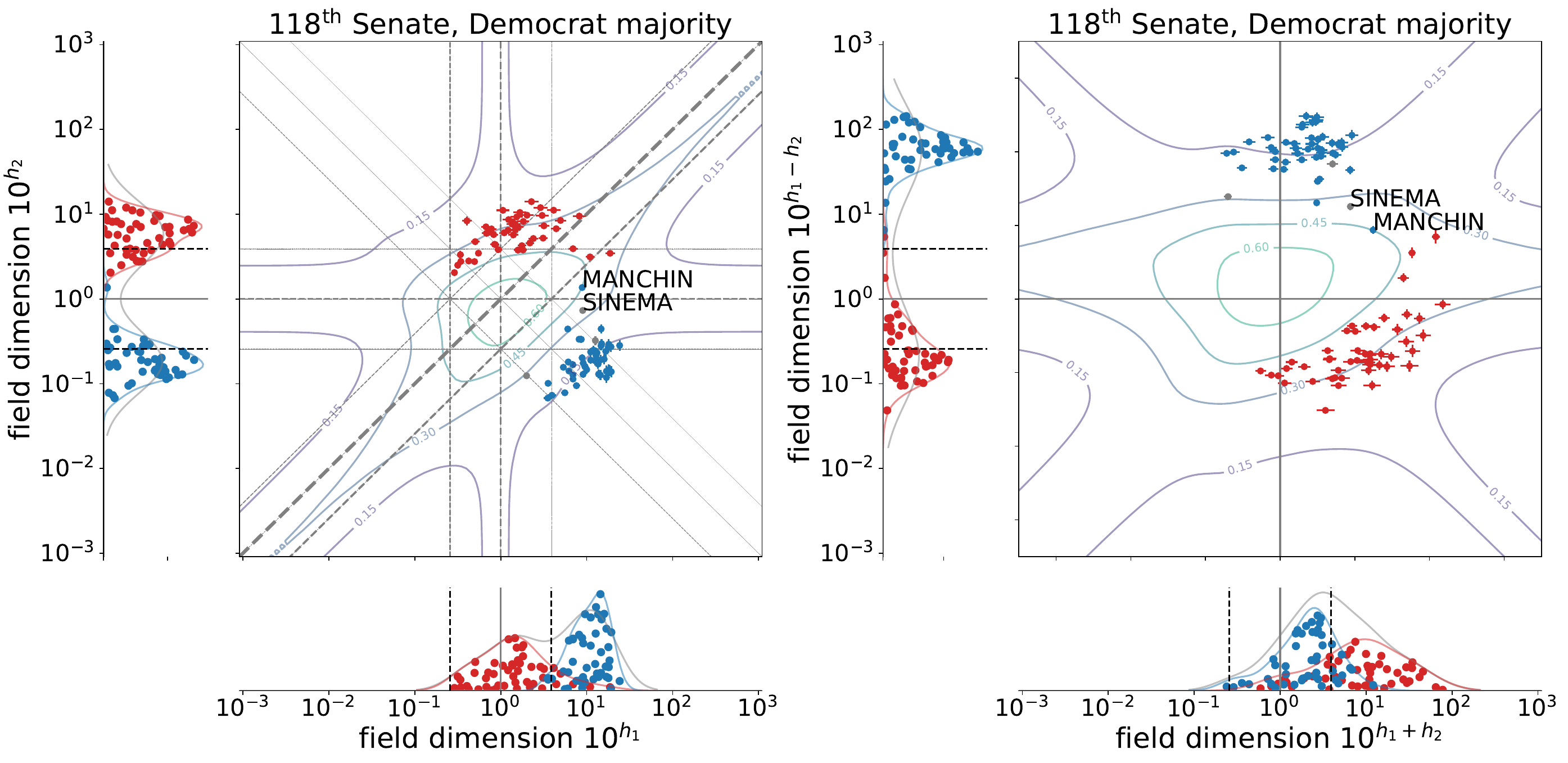}\includegraphics[width=.49\linewidth]{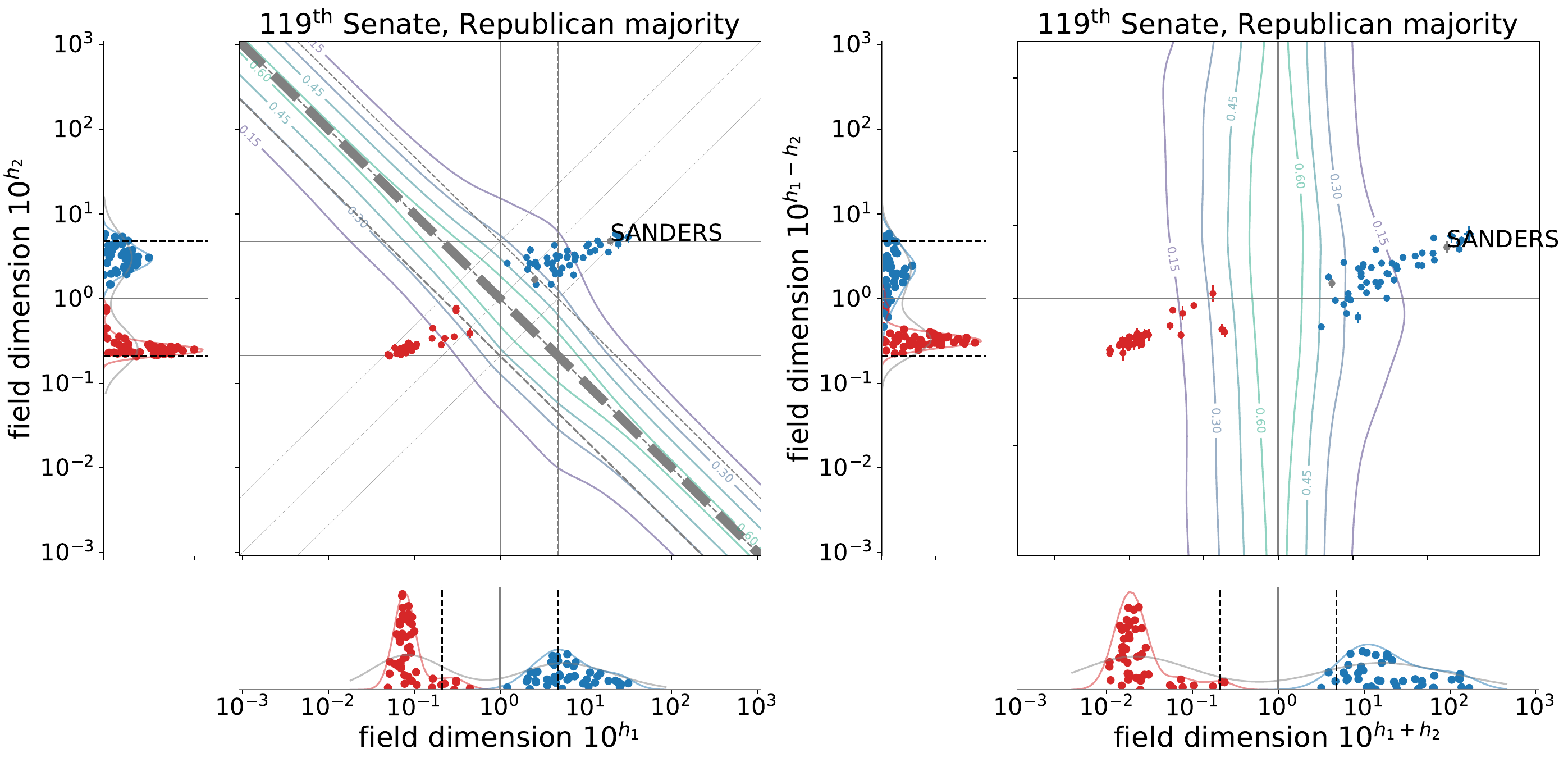}\\
	\caption{Parameter projection for $K=3$. \session{113} through \session{119} sessions, excluding the \session{117}.}\label{fig:maps4}
\end{figure*}

\end{document}